\documentclass[final,3p]{elsarticle}
\usepackage{lineno,booktabs,amsmath,graphicx,url}
\usepackage[hidelinks]{hyperref}
\usepackage{tikz}
\usetikzlibrary{positioning,arrows.meta,fit,backgrounds}
\graphicspath{{fig/}}
\journal{Sustainable Cities and Society}

\newcommand{\AudN}{122}

\newcommand{\AudFieldN}{40}
\newcommand{\AudFieldValue}{39}

\newcommand{\AudFieldGfa}{25}
\newcommand{\AudFieldGfaMissed}{15}

\newcommand{\ScrReleasedA}{112}

\newcommand{\ScrAuditA}{20}

\newcommand{\ScrPdfHitYA}{100\%}
\newcommand{\ScrPdfHitNA}{100\%}

\newcommand{\ScrMktN}{40}

\newcommand{\GtwoPrecision}{68.4\%}
\newcommand{\GtwoExposure}{82.5\%}
\newcommand{\GtwoNoneNamed}{9}
\newcommand{\GtwoNoneNamedN}{13}
\newcommand{\GtwoHq}{2}

\newcommand{\BrOwners}{101}
\newcommand{\BrNamed}{87}
\newcommand{\BrExposed}{32}

\newcommand{\BrStatusKnown}{30}
\newcommand{\BrMetricMatch}{14}
\newcommand{\BrReadable}{0}
\newcommand{\BrJuris}{11}
\newcommand{\ExpoShare}{43.6\%}
\newcommand{\ExpoLo}{28.7\%}
\newcommand{\ExpoHi}{60.2\%}

\newcommand{\GoneAsHone}{95.1\%}

\newcommand{\GoneSentence}{recall of the non-disclosing classification was 95.1\% (firm-clustered 95\% interval 90.2\% to 99.0\%; 6 of 122 reports), against a bar of 90\%}
\newcommand{\BrStatus}{provisional: the machine screen and blind adjudication registered for it are in progress, and the final counts replace these when that gate closes}
\newcommand{\GtwoSentence}{Judged against the final hand codes, the extraction failed all four registered criteria}\begin{document}
\begin{frontmatter}

\title{Printed but not benchmarkable: most building-decarbonisation
disclosure cannot be matched to the pathways that stranding regulation
assumes}

\author[yale]{Jingyi Xu\corref{cor}}
\ead{jy.xu@yale.edu}
\cortext[cor]{Corresponding author.}
\author[um1,um2]{Minghui Cheng}
\ead{mxc3063@miami.edu}
\author[goog]{Anchen Sun}
\ead{anchensun@google.com}
\affiliation[yale]{organization={Yale Center for Ecosystems in Architecture (Yale CEA), Yale University}, country={USA}}
\affiliation[um1]{organization={Department of Civil \& Architectural Engineering, University of Miami}, city={Coral Gables}, postcode={33146}, state={FL}, country={USA}}
\affiliation[um2]{organization={School of Architecture, University of Miami}, city={Coral Gables}, postcode={33146}, state={FL}, country={USA}}
\affiliation[goog]{organization={Google}, country={USA}}

\begin{abstract}
Cities are beginning to enforce carbon limits on existing buildings. Science-based decarbonisation pathways set those limits one asset type and one jurisdiction at a time. Owners, however, report for the whole firm. We measure what that mismatch costs on two sets of public corporate reports: a census of 502 reports from the 119 listed built-environment firms with a collected report inside a 2{,}246-firm panel (2003--2023), and 519 real-estate reports from 101 firms (2007--2024). BeDA, a multimodal language-model tool whose reliability we test first, read them. Running the pathway frameworks' own entry tests over published disclosure: $16.5\%$ of census reports ($43.7\%$ of real-estate reports) print an operational carbon intensity per square metre; $6.6\%$ ($25.0\%$) can be matched to a pathway for their property type in a covered jurisdiction; and only $5.0\%$ ($16.4\%$) disclose the floor area they divided by. Of the failures at the pathway test, $82$--$84\%$ follow from reports lumping the portfolio together and $16$--$18\%$ from a missing curve in the pathway library. The obstacle is the reporting unit, not missing data. The rate is roughly
twice as high for European as for US listings ($65$--$71\%$ versus
$35\%$ in listed real estate). We also show that a US portfolio's carbon verdict cannot be worked out from disclosure at all. Within one climate zone, the pathway's carbon limit varies by up to $2.79\times$ with the electricity subregion, which no report names; its energy limit does not move. Extraction is checked against the source PDFs ($97.3\%$ of extracted intensities appear verbatim) and repeats on a second extractor ($\kappa{=}0.97$). Recall of the non-disclosing class was \GoneAsHone{} in a blinded hand audit of \AudN{} reports. The fix follows from the measurement: split intensity by asset type and jurisdiction, and report floor area.
\end{abstract}

\end{frontmatter}

\section{Introduction}
Buildings and construction took 28\% of global energy use in 2024 and caused around 37\% of global carbon emissions. Running buildings alone emitted 9.9\,GtCO$_2$ in 2024: about 26\% of all energy-related CO$_2$, and 3.5\,GtCO$_2$ more than the IEA net-zero scenario allows \citep{unepgsr2026}. That makes the sector the single largest decarbonisation lever cities have. Progress depends not only on physically retrofitting the building stock \citep{berrill2022residential,ang2023eightcities}. It also depends on whether the \emph{disclosed} data that steer investment and regulatory attention can support the comparison the regulation assumes. Building
performance mandates are proliferating and tightening: sixteen US
state and local governments had enacted a building performance
standard by December 2025 \citep{imt2026matrix}, and the National
BPS Coalition counts ``49 governments, including 7 states and 42
localities'' as members \citep{nbpsc2026}; New York City's Local Law~97 caps carbon emissions per unit of floor area for large buildings, by occupancy group \citep{nyc2019ll97,nycdob2026rule}; 93\% of covered properties filed their first compliance reports by April 2026 \citep{nycdob2026filing}; Boston's BERDO admits a
``Building Portfolio'' as a compliance unit while retaining
per-building reporting \citep{boston2021berdo}; state standards in
Maryland, Colorado and Oregon extend per-square-foot carbon or energy
intensity limits to commercial buildings of 35,000, 50,000 and
35,000~ft$^2$ and above
\citep{md2022csna,mdcomar2628,co2021hb1286,coreg28,or2023hb3409}; the EPBD recast
mandates national per-m$^2$ trajectories \citep{eu2024epbd}, although
all 27 Member States missed its transposition deadline and the
Commission opened infringement procedures in July 2026
\citep{ec2026infringement}, so the building-level databases that would
close the data gap are delayed EU-wide. Company-level disclosure rules, meanwhile, are shrinking. The CSRD Omnibus drops roughly $90\%$ of previously covered companies \citep{eu2026omnibus}, and the SEC has proposed withdrawing its climate-disclosure rules \citep{sec2026rescission}. The Californian substitute is slipping: the Ninth Circuit has blocked SB~261 climate-risk reporting pending appeal since 18 November 2025, and CARB proposed in a 15-day notice of 27 July 2026 to push the first SB~253 Scope~1--2 reporting deadline from 10 August to 10 November 2026 \citep{carb2025sb261,carb2026notice}. Both movements sharpen the same
question: what the corporate reports that \emph{do} exist can support.
Cities enforce building caps with required benchmarking data, an information tool shown to change building energy use \citep{kontokosta2020audits}. The legal literature on standard design has itself noted the thin evidence base: ``[a]s of this writing, there is little rigorous analysis of existing BPSs to provide policymakers with an evidentiary basis for making these design choices'' \citep[p.~10277]{spiegelfeld2022building}. Its catalogue of compliance-flexibility mechanisms recorded, as of 2022, only Boston and Washington, D.C.\ letting ``[o]wners of multiple buildings \ldots\ demonstrate compliance on a portfolio basis'' \citep[Table~1, p.~10358]{spiegelfeld2022tradable} --- the point at which the owner's total, rather than each building, becomes the compliance unit. Above the individual building, portfolio-level oversight must rely on thousands of self-reported firm disclosures --- oversight by owners screening transition risk, by investors, and by supervisors who have found that ``property-level information on emissions or energy usage \ldots\ is not currently collected or systematically available'' \citep{fed2024pilot}.

Prior work on LEED certification adoption across top architecture firms
\citep{xu2025leed} showed both the feasibility and the policy value of
large-scale assessment of built-environment sustainability. That study,
however, rested on a third-party-\emph{verified} signal (certification)
covering only a minority of activity. The decarbonisation commitments
regulators must now act on are overwhelmingly \emph{self-disclosed and
unverified}, and the first quantitative evaluations of large corporate
climate initiatives find mixed progress against their own pledges
\citep{ruizmanuel2023initiatives}. Credibility cannot be assumed; it
must be established. BeDA (the Built-environment Decarbonisation-disclosure
Auditor) takes that step, moving from analysing a verified standard to
building a reliable instrument where none exists.

The deficiency is one of method. State-of-practice detectors such as
ClimateBERT \citep{webersinke2022climatebert,bingler2022cheap} and the
Loughran--McDonald lexicon
\citep{loughran2011liability} share three limitations. First, each comes from a single model with \emph{no cross-model reliability evidence}. A score that changes when you swap the model underneath cannot support a regulatory decision. Second, they are economy-wide or
finance-framed and not built-environment-specific at panel scale. Third,
they are not tested against the \emph{physical-data requirements} of
stranding frameworks \citep{crrem2023,pcaf2023guidance}: a credibility
score is only actionable for stranding if the underlying disclosure can
be mapped to a floor-area-normalised intensity. Large language models (LLMs) can become \emph{less} reliable as they get bigger \citep{zhou2024reliable}, and LLM-based measurement is now known to need systematic validation \citep{pnas2025wellbeing}. For built-environment disclosure, neither reliability across model families nor physical-data fitness has been measured.

Nobody has yet measured whether public corporate disclosure can feed a science-based pathway comparison \emph{from disclosure documents on a defined frame, carried through to pathway applicability}. Industry has measured pieces of it. FTSE Russell reports that 10\% of the 40 largest listed real-estate firms could not produce a comparable per-floorspace emissions metric \citep{ftserussell2024}. KPMG/EPRA count, sector by sector, which of 120 (2023 edition) and 110 (December 2025 edition) European listed firms disclose kWh/m$^2$ and kgCO$_2$/m$^2$ \citep{kpmgepra2023,kpmgepra2025}.
What no prior work does is run the pathway frameworks' \emph{own admission gates} over public disclosure, gate by gate, with the failures decomposed. The gates require an operational carbon intensity printed per square metre, resolving to a covered jurisdiction, a property type, a single named curve, with the floor area it was divided by. The gates are not ours: the reference guide of CRREM's Risk
Assessment Tool excludes assets missing mandatory inputs from its
results, and GRESB's import rules refuse assets in uncovered countries
or unsupported property types \citep{crrem2023,gresbcrremtool}. CRREM retired that Excel tool on 1 July 2026 and replaced it with an open-access dataset and calculation library, saying its pathways, methodology and underlying datasets are unchanged \citep{crrem2026retire}. So the admission logic we test against is that of the CRREM Global Pathways V2.04 dataset and its open-access successor, not of a spreadsheet. Delivering that measurement is
the \emph{primary objective} of this study. Two \emph{secondary
objectives} support it: profiling the stranding-readiness of the
disclosing real-estate stratum, and establishing the reliability of
the enabling LLM instrument before either measurement is used. These
objectives yield three contributions:
\begin{enumerate}
\item \textbf{The constructibility cascade.} On a 502-report census of
listed built-environment firms and a 519-report real-estate corpus,
$16.5\%$ and $43.7\%$ of firm-reports print an operational carbon
intensity per m$^2$; $6.6\%$ and $25.0\%$ reach a
property-type-specific pathway in a covered jurisdiction; $3.0\%$ and
$10.6\%$ reach an exact country$\times$property-type curve; and
$5.0\%$ and $16.4\%$ disclose the floor-area denominator itself ---
in the census, \emph{fewer} reports disclose their own denominator
than can be matched to a pathway. Failures at the pathway gate split into reports that lump the portfolio together ($82$--$84\%$) and curves missing from the pathway library ($16$--$18\%$). The European rate is roughly twice the US rate
($65$--$71\%$ versus $35\%$ in listed real estate,
Section~\ref{sec:readiness}).
\item \textbf{A non-identification result for US portfolios.} Within a fixed (property type $\times$ climate zone) cell, CRREM's V2.04 energy limit is identical in every electricity subregion, while its carbon limit varies by up to $2.79\times$ \citep{lbnl2024crrem}. No report in either corpus names its subregion, so a US portfolio's carbon verdict cannot be built from disclosure even where the intensity is printed, though its energy verdict could be (Section~\ref{sec:grid}). A disclosure-derived stranding screen on the
constructible real-estate reports is reported with this band attached.
\item \textbf{The credentialed instrument:} the spine scores are
reused unmodified from the SSSR corpus; BeDA's contribution is their
credentialing and application. The reliability tests show the score is stable across models ($r=0.845$, $n=407$) and across model \emph{families} ($r{\approx}0.74$--$0.82$). It agrees with three independent outside criteria: LSEG-ESG ratings, RepRisk incident counts, and Science Based Targets initiative (SBTi) target validation. We report in full the one place it differs from human coders (Appendices). Extraction is checked against the source (97.3\% of extracted intensities appear verbatim in the source PDF), which floors hallucination, and repeats on a second extractor ($\kappa{=}0.97$) and on three independent non-Google vendors ($\kappa{=}0.86$--$0.89$), which bounds extractor-specific and shared-provider error. Recall of the non-disclosing class was \GoneAsHone{} in a blinded hand audit of \AudN{} reports (Section~\ref{sec:limitations}).
\end{enumerate}

This is ultimately a question of urban governance. Cities are the
level at which building-stock decarbonisation is regulated, through
performance mandates such as Local Law~84/97 and the EPBD recast. Those
mandates can be enforced only against what can be measured on the
floor-area carbon-intensity basis the pathway uses. Building-level compliance is
assessed through benchmarking submissions. At the portfolio level,
where owners, investors and entity-level disclosure mandates operate,
the corporate firm-report is the most scalable carrier of
building-stock carbon data. Our object of study is the
\emph{governability} of the urban building stock under these regimes.
Our result is an urban-policy diagnosis: a measurable,
jurisdiction-specific gap between the disclosure this portfolio-level
oversight receives and the disclosure the stranding frameworks
presuppose. The LLM machinery delivers that diagnosis at panel scale; its role is
instrumentation.

The remainder is organised as follows. Section~2 reviews related work.
Section~3 describes the data and the BeDA instrument. Section~4 reports
the disclosure-fitness gap and stranding-readiness profile; instrument
reliability and validity are summarised there, with the leakage probe,
the LLM-versus-human measurement study, baseline comparisons, and CRREM
construction details in the appendices. Sections~5--6 discuss mechanism,
limitations, and policy.

\section{Literature review}
Four strands of prior work bear on the credible, scalable auditing of
building-decarbonisation disclosure. Each contributes part of what a
regulatory instrument needs. This study sets out to assemble what is
still missing: a single built-environment-specific instrument whose
reliability and confound-controlled validity are both established. Table~\ref{tab:landscape} summarises how the present study
builds on these strands and closes that remaining gap.

\subsection{Text-based disclosure-quality detection}
Text classifiers dominate automated disclosure assessment. ClimateBERT
\citep{bingler2022cheap} grades climate-disclosure paragraphs (under the
Task Force on Climate-related Financial Disclosures, TCFD, framework) at
high accuracy and flags a large share of them as low in specificity
(what its authors term ``cheap talk''). The Loughran--McDonald lexicon
\citep{loughran2011liability} remains the standard dictionary-based sentiment
benchmark. Carbon-disclosure intensity has further been linked to firm
idiosyncratic volatility \citep{perera2023carbon}, confirming that disclosure
quality carries real financial consequences. These methods are valuable, but
three properties limit their use for regulation: each operates from a single
model or lexicon, none reports cross-model reliability, and none is
built-environment-specific. Regulatory triage therefore needs evidence that the score survives a
change of model. We evaluate
both as baselines on the same set of confound-controlled validity tests
(Section~\ref{sec:baselines}, \ref{app:validity}).

\subsection{Greenwashing detection with machine learning and large language models}
A growing literature applies machine learning and large language models
to detect firm-level greenwashing
\citep{eneco2025greenwashing,zhang2024greenwashingai}. Related evidence
shows how regulatory shocks reshape greenwashing behaviour
\citep{zhang2022greenwashingreg} and links ESG disclosure to corporate
financial irregularities \citep{yuan2022esgirregular}. Validation
typically rests on accuracy/F1 and on correlation with external ESG ratings,
whose reliability as a criterion is itself contested
\citep{berg2022aggregate,jbr2025greenwashing}. However, such correlations are reported as validity without controlling
for confounds that inflate both the score and the criterion, above all
disclosure volume. The confound-controlled design of this study
(Section~\ref{sec:baselines}) adjusts for disclosure volume directly
and for firm composition (sector, listing region, reporting year)
categorically.

\subsection{Large language models as measurement instruments: reliability and validity}
Using an LLM as a measurement instrument calls for the same reliability
and validity checks that certify any measurement tool in the social
sciences (psychometric validation
\citep{serapiogarcia2025psychometric,demszky2023llmpsych}; annotation
reliability against human coders
\citep{gilardi2023chatgpt,ziems2024css}). Without such checks, LLMs can
systematically misestimate \citep{pnas2025wellbeing,xie2026llmrealism},
larger instruction-tuned models can be \emph{less} reliable
\citep{zhou2024reliable}, and pre-trained models carry measurable
normative biases \citep{schramowski2022biases}. Sustainability reports
are inherently multimodal, and capable multimodal document models are
now well established in the machine-learning literature
\citep{gemini2023,gemma2025}. That
literature, however, validates them for task accuracy, which is a
different requirement from the psychometric reliability a regulatory
instrument needs. For sustainability disclosure specifically,
cross-\emph{family} reliability, pretraining-reputation leakage, and
confound-controlled validity remain unaddressed.

\subsection{Built-environment decarbonisation standards}
CRREM defines science-based 1.5--2\,$^\circ$C carbon- and energy-intensity
pathways by property type and country \citep{crrem2023,crrem2025pathways}, and the
PCAF--CRREM--GRESB guidance standardises real-estate GHG accounting
\citep{pcaf2023guidance}. The SBTi Buildings Sector Criteria
(Version~1.2, effective 20 July 2026) carry the same admission
structure into corporate target-setting: intensity targets must be
expressed in kgCO$_2$e/m$^2$, with one floor-area definition applied
consistently across the GHG inventory and the base and target years
(criterion Buildings-C10), and buildings outside the typologies and
geographies covered by the SBTi--CRREM in-use operational pathways
must fall back to an `Other' pathway (Buildings-C5), only 29\% of
those pathways having been pilot-tested against real-world data
\citep{sbti2026buildings}. Mispriced transition risk translates into large
stranded-asset losses \citep{semieniuk2022stranded}, with owner exposure
concentrated enough to trigger climate-policy resistance
\citep{vondulong2023stranded}. Credible stranding assessment must
engage the legal frame \citep{wetzer2024climatelaw}. Certification schemes
such as LEED provide a third-party-\emph{verified} signal whose firm-level
adoption has been analysed at scale \citep{xu2025leed}; the present study is
complementary, addressing the un-certified, self-disclosed majority.
However, all of these \emph{require} floor-area (per-m$^2$) normalisation,
and none audits whether disclosure actually supplies it. The closest
precedents in ambition are panel-scale, data-driven urban measurements
for decision support: interpretable deep-learning studies of urban
carbon \citep{zhang2024carbon275,zhang2025digitalai} and
urban-infrastructure forecasting \citep{xia2024water}. These are
numerical-forecasting methods, however, and thus methodologically
distinct from the document auditing developed here.

\subsection{The gap is documented; its size is not measured}
The data gap itself is well attested, from three directions, none of
which measures it from disclosure documents. Standard-setters and
supervisors assert it: CRREM's own authors report by survey that only
6\% of financial institutions had complete asset-level information
\citep{unepfi2022}, the ECB finds fewer than ten supervised banks hold
majority-actual energy-performance data on real-estate collateral
\citep{ecb2024cre}, and the US Federal Reserve's pilot climate
exercise states that property-level emissions and energy data ``is not
currently collected or systematically available'' \citep{fed2024pilot}.
Supervisory demand for the same datum is being codified rather than
relaxed: in the EBA's final draft Pillar~3 ITS of 22 June 2026 --- a
simplification round under which large institutions ``will disclose
37\% less datapoints than now'' --- the energy-performance template
for loans collateralised by immovable property survives as Template
EU~CRFR3, the EBA stating that ``including EPC data in Template~2
remains important'', and it gains a column for exposures whose
collateral energy-performance data ``are unavailable and can not be
estimated'' \citep{eba2026its}.
Industry counts disclosure at the top of the funnel: FTSE Russell for
the 40 largest listed real-estate firms \citep{ftserussell2024},
KPMG/EPRA per property sector for 120 European firms in the 2023
edition and for 110 EPRA-award companies over 2022--2024 in the fifth
(December 2025) edition \citep{kpmgepra2023,kpmgepra2025}, EPRA
itself for GHG and net-zero-target disclosure across 150 listed
European members as of 30 June 2025 --- 91\% report GHG emissions,
an analysis of reporting scope and target-setting rather than of
intensity \citep{epra2025ghg} ---, and the Global Real Estate Engagement Network reports that the share
of the Top-120 FTSE EPRA Nareit Developed constituents disclosing
``CRREM or a similar transition risk analysis at the portfolio
level'' rose from 7\% in 2024 to 11\% in 2025, while ``many still do
not provide access to this type of data'' \citep{green2025}. GREEN's
indicator is an \emph{outcome}: whether the company itself publishes
a completed analysis. The cascade below measures the complementary
\emph{input} condition --- whether an outsider could construct such
an analysis from the disclosure at all --- gate by gate, with the
failure decomposed by gate, so the two figures answer different
questions and are not expected to coincide. Commercial providers route
\emph{around} disclosure: the Measurabl--CRREM company-level dataset
substitutes machine-learning estimates trained on proprietary meter
data, citing ``limited visibility into the underlying asset details''
\citep{measurabl2026}; academic assessments of transition plans
likewise bypass reporting via external asset inventories in sectors
that have them \citep{kampmann2026,saleh2026} --- an option the
built environment lacks, since no global asset tracker carries in-use
intensity, floor area and jurisdiction for buildings. Corporate
disclosure audits at high-impact venues measure completeness,
consistency and target integrity
\citep{klaassen2021,cohen2026revisions,jiang2025targets,bjorn2023sbt,jouffray2025ocean},
not fitness-for-use against a named decision algorithm's admission
rules. What is missing is the joined measurement: the gate-by-gate
rate at which \emph{public disclosure} clears the pathway frameworks'
own admission requirements, with the failures decomposed and the
denominator disclosure itself measured. That is what this study
supplies, and the enabling instrument is credentialed before it is
used (Section~\ref{sec:instrument}).

A reader of the US trade statistics may object that the gap looks
smaller from that side: Nareit's 2026 industry report finds that
87\% of the top-100 US equity REITs report an energy-use intensity
and 95\% report Scope~1--2 emissions \citep{nareit2026}. Four
differences reconcile the two pictures. The Nareit metric is
\emph{energy}, not carbon: the report's floor-area quantity is energy
per square foot, while the carbon-intensity figure it prints is
denominated per million dollars of enterprise value (EVIC), not per
unit floor area. Its frame is a curated top-100 by market
capitalisation (98\% of index market cap), not a census of listed
built-environment firms. Its rates are expressed as a percentage of
market capitalisation rather than of firms. And it counts whether a
metric is reported, not whether the reported metric can be carried
to a pathway --- disaggregated by property type and jurisdiction,
with the floor area it divides by. The cascade below measures that
last condition on a census frame, and its constructibility rates are
conditional on the very disclosure that Nareit counts.

\begin{table}[!htbp]\centering
\caption{Capabilities required of a regulatory credibility instrument, and
their coverage by prior approaches versus the present study (BeDA).
``spec.''~$=$~specified by the standard; industry has measured the
top-of-funnel disclosure rate
\citep{ftserussell2024,kpmgepra2023,kpmgepra2025,green2025,nareit2026}, but the gate-by-gate pathway
constructibility and the floor-area-denominator rate remain
unmeasured --- the denominator row is the one this table's argument
rests on.}
\label{tab:landscape}
\begin{tabular}{lcccccc}
\toprule
 & Built-env & Cross- & Cross- & Leakage & Confound- & Per-m$^2$\\
Approach & specific & model & family & tested & ctrl.\ valid. & constructibility\\
\midrule
Text-based detection      & no  & no   & no  & no   & no      & no\\
ML/LLM greenwashing       & no  & no   & no  & no   & partial & no\\
LLM-as-measurement        & no  & some & no  & rare & varies  & no\\
Built-env standards       & yes & n.a. & n.a.& n.a. & n.a.    & spec.\\
Industry disclosure counts& yes & n.a. & n.a.& n.a. & n.a.    & top-of-funnel\\
\textbf{BeDA}             & \textbf{yes} & \textbf{yes} & \textbf{yes} & \textbf{yes} & \textbf{yes} & \textbf{yes}\\
\bottomrule
\end{tabular}
\end{table}

\section{Methodology}

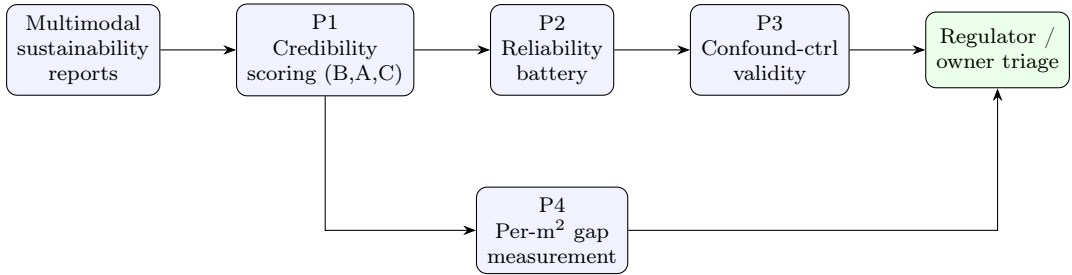
\begin{figure}[!htbp]\centering
\begin{tikzpicture}[node distance=6mm and 10mm,
  box/.style={draw,rounded corners,align=center,font=\footnotesize,
  inner sep=4pt,minimum height=10mm,fill=blue!5},
  >={Stealth[]}]
\node[box] (rep) {Multimodal\\sustainability\\reports};
\node[box,right=of rep] (p1) {P1\\Credibility\\scoring (B,A,C)};
\node[box,right=of p1] (p2) {P2\\Reliability\\battery};
\node[box,right=of p2] (p3) {P3\\Confound-ctrl\\validity};
\node[box,below=12mm of p2] (p4) {P4\\Per-m$^2$ gap\\measurement};
\node[box,right=of p3,fill=green!8] (out) {Regulator /\\owner triage};
\draw[->] (rep)--(p1); \draw[->] (p1)--(p2); \draw[->] (p2)--(p3);
\draw[->] (p3)--(out); \draw[->] (p1) |- (p4); \draw[->] (p4) -| (out);
\end{tikzpicture}
\caption{The BeDA instrument. P1 produces the credibility scores; P2--P3
establish that those scores are \emph{reliable} (stable when the scoring
model is changed) and \emph{valid} (aligned with independent external
benchmarks once confounding factors are held constant); P4 applies the
validated instrument to measure the disclosure-fitness gap.}
\label{fig:pipeline}
\end{figure}

Fig.~\ref{fig:pipeline} gives an overview of the four-phase BeDA
pipeline developed in this section.

\subsection{The SSSR disclosure corpus}
This study builds on \emph{SSSR} (Smart Service Social Responsibility),
a framework for scoring corporate sustainability reports that was
developed in two stages. First, a systematic literature review of 264
articles on responsibility in smart-service ecosystems established the
conceptual structure: five responsibility dimensions (environmental,
social, ethical, legal, economic), two spaces (physical and virtual), and
eight stakeholder groups \citep{nan2025governance}. Second, this
structure was operationalised at panel scale. The published SSSR
analyses use a cross-sector hybrid pipeline that combines TF-IDF
keyword analysis with LLM-based semantic extraction, built on LLaMA 3.2
variants including a vision-capable model that reads charts, tables,
and infographics directly from report PDFs \citep{xia2025global}. The fixed-rubric dimension scores that the
present study reuses were produced for the SSSR corpus with
\texttt{gemini-3.1-pro-preview} under the rubric of
Table~\ref{tab:rubric} (Section~\ref{sec:instrument}). The published
SSSR analyses cover 7{,}858 sustainability reports from five
smart-service industries (technology, financial, services, healthcare,
and consumer goods) over 2000--2023. The underlying report collection
is broader: the fixed-rubric score panel reused here covers 2{,}246
publicly listed firms across sectors (9{,}166 firm-years, 2003--2023),
including the built-environment industry divisions this study audits
(Section~\ref{sec:scope}). Corpus-scale LLM extraction from
sustainability reports is an established method --- Forster et al.\
process 9{,}173 PDFs of the STOXX Europe 600 with explicit unit
standardisation \citep{forster2026}, and public benchmarks for GHG
extraction now exist \citep{beck2025} --- so the pipeline is a methods
statement here, not a contribution. BeDA's specialisation is the
built-environment credentialing (the reliability battery, leakage
probe, and confound-controlled validity of
Section~\ref{sec:reliability}) and one design point offered with a
rationale rather than as a demonstrated improvement: whole-PDF
long-context reading instead of chunk-and-retrieve, avoiding the
retrieval-recall error channel; no head-to-head comparison exists.

For the present study, the SSSR rubric assigns each report three groups
of scores, which the parent framework labels as dimensions A, B, and C
(Table~\ref{tab:rubric}). Because these letter labels are specific to
SSSR and not established conventions in the built-environment literature,
we refer to them by their descriptive names throughout:
\begin{itemize}
\item \textbf{Topic breadth (SSSR-A, 0--40):} how broadly a report
covers ESG themes across physical and virtual spaces.
\item \textbf{Standards compliance (SSSR-B, 0--32):} the degree to
which a report aligns with recognised disclosure standards such as the
Global Reporting Initiative (GRI), the Sustainability Accounting
Standards Board (SASB), TCFD, and third-party assurance frameworks. This is the dimension
BeDA certifies and builds its credibility audit on. We call it the
\emph{standards-compliance spine}, or simply the \emph{spine}, because
it is the structural backbone of the audit: it is the one score that
proves stable across models and model families
(Section~\ref{sec:reliability}).
\item \textbf{Multimodal evidence (SSSR-C):} three sub-scores capturing
visual evidence (C1, 0--10), tabular transparency (C2, 0--10), and
cross-modal consistency between text and visuals (C4, 0--10).
\end{itemize}
The topic-breadth and multimodal scores are retained for the reliability
analysis (Section~\ref{sec:reliability}) and shown there to be
substantially less stable than the spine.

\begin{figure}[!htbp]\centering
\includegraphics[width=.92\linewidth]{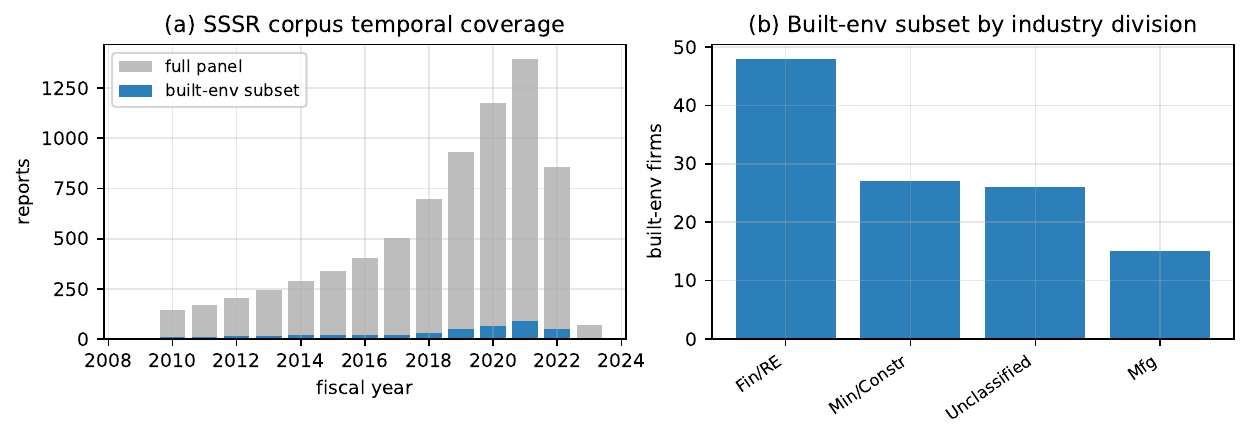}
\caption{The SSSR corpus. (a) Temporal coverage, full panel vs the
built-environment subset, 2003--2023. (b) Built-environment subset by
industry division.}
\label{fig:corpus}
\end{figure}

The score panel used in this study spans 2003--2023
(Fig.~\ref{fig:corpus}a); its built-environment subset covers the
construction, real-estate, and building-materials industry divisions
that govern building decarbonisation (Fig.~\ref{fig:corpus}b). Because SSSR scores are
LLM-produced, establishing their reliability is itself a primary
contribution of this paper.

\begin{table}[!htbp]\centering
\caption{The SSSR scoring rubric. Standards compliance (SSSR-B) is the
spine used by BeDA. The parent rubric's sub-score C3 is not used in this
study; the original dimension labels are retained for comparability with
the parent corpus \citep{xia2025global}.}
\label{tab:rubric}
\begin{tabular}{llp{0.5\linewidth}}
\toprule
SSSR dim. & Max & What it measures\\
\midrule
A (topic breadth)   & 40 & Coverage of ESG dimensions across physical and virtual spaces\\
B (standards compliance)   & 32 & Alignment with recognised disclosure standards (the \emph{spine})\\
C1 (visual evidence)  & 10 & Substantive figures and imagery\\
C2 (tabular transparency)  & 10 & Quantitative tables\\
C4 (cross-modal consistency)  & 10 & Agreement between text and visuals\\
\bottomrule
\end{tabular}
\end{table}

\subsection{Study scope and data}\label{sec:scope}
\begin{figure}[!htbp]\centering
\includegraphics[width=.6\linewidth]{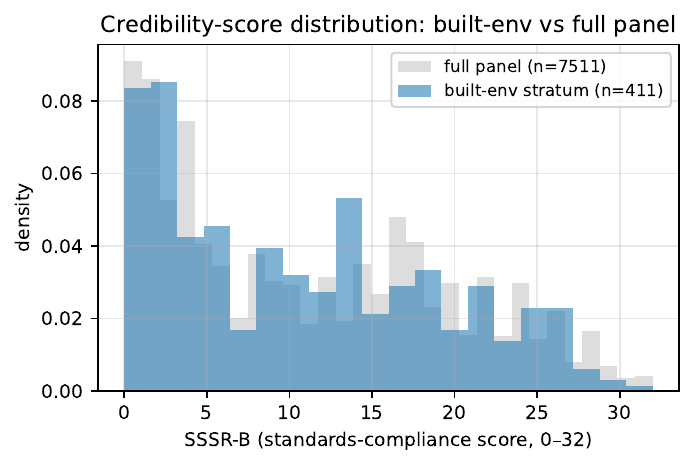}
\caption{Credibility-score distribution: the built-environment stratum vs
the full panel.}
\label{fig:bdist}
\end{figure}

The unit of analysis is the built-environment firm-report. The design
is an observational, cross-sectional audit of a retrospective
firm-report panel; no intervention is involved, and inferential
analyses pool firm-year observations with firm-clustered standard
errors. A global firm panel (2{,}246 firms; 9{,}166 firm-years;
2003--2023) of multimodal sustainability reports is reused as one
dataset, following the panel-as-dataset design established in prior
interpretable deep-learning studies of urban carbon
\citep{zhang2024carbon275,zhang2025digitalai}. An industry crosswalk
resolves each firm to a built-environment flag via SEC-SIC with a global
ticker fallback, in three states: built-environment, not
built-environment, or unresolved. Unresolved firms are carried explicitly, since reassigning them would
bias the gap statistic downward.
This yields the sample of Table~\ref{tab:sample}: a built-environment
lead stratum of 116 firms on the core SIC scope (71 outcome-anchored
against an independent incident database), 127 tagged firms once the
building-services expansion below is added; 119 of the 127 have at
least one collected report, and every collected report of those 119
firms (502 reports) is the census of Section~\ref{sec:gap} --- the
remaining 8 tagged firms have no PDF in the scored corpus. Unresolved
firms (21.6\% of the panel) are handled by rate-extrapolation (defined
in Section~\ref{sec:implementation}). The
full panel is retained as the comparison frame that carries inferential
weight. Fig.~\ref{fig:bdist} contrasts the
credibility-score distribution of the built-environment stratum against the
full panel.

\begin{table}[!htbp]\centering
\caption{Sample composition (built-environment industry crosswalk).}
\label{tab:sample}
\begin{tabular}{lr}
\toprule
Quantity & Value\\
\midrule
Panel firms & 2{,}246\\
Panel firm-years (2003--2023) & 9{,}166\\
Built-environment lead stratum (core SIC scope) & 116\\
\quad expanded scope (+11 building-services firms) & 127\\
\quad with $\geq$1 collected report (census firms) & 119\\
\quad outcome-anchored (incident DB) & 71\\
Crosswalk-unresolved (rate-extrapolated) & 21.6\%\\
Built-env reports cross-scored (reliability) & 407\\
\bottomrule
\end{tabular}
\end{table}

\subsubsection*{Scope alignment with the IPCC AR6 Ch.~9 / IEA Buildings perimeter}
The crosswalk's narrow operational definition is SEC-SIC-anchored:
SIC 1500--1799 (construction), 2400--2499 (lumber and wood products),
3240 (cement), 3250--3299 (clay, concrete, gypsum and stone building
materials), 3420--3460 (building hardware and metalwork), 6500--6599
(real estate), and 6798 (REITs). These are the firms whose
disclosures are most directly addressed by CRREM and GRESB. The IPCC
AR6 WGIII Ch.~9 \citep{cabeza2022buildings} and the IEA/GlobalABC
Tracking Buildings perimeter \citep{iea2024buildings},
however, additionally include the building-services equipment supply
chain (HVAC, lighting, residential and commercial environmental
controls) and building-materials retail. These firms' product decisions
govern operational energy-use intensity (EUI) per m$^2$ in the existing
and future stock. We therefore audit the narrow scope against an
architecturally aligned expansion that adds, conservatively, SIC 3585
(HVAC manufacturing), 3640--3648 (lighting fixtures), 3822 (automatic
environmental controls; excluding 3823--3829, laboratory and scientific
instruments), and 5210--5211 (building-materials retail), plus the
Yahoo Finance industry classification ``Home Improvement Retail''. The
expansion adds
eleven firms (Trane, Carrier, Lennox, Johnson Controls, AAON, Acuity
Brands, LSI Industries, Home Depot, Lowe's, Builders FirstSource,
Kingfisher). It takes the built-environment lead stratum from
$n{=}116$ to $n{=}127$ tagged firms ($+9.5$\%); 119 of the 127 have at
least one collected report and constitute the 502-report census of
Section~\ref{sec:gap}, the other 8 having no PDF in the scored corpus. The reliability and
controlled-validity evidence
(Sections~\ref{sec:reliability}--\ref{sec:validity}) is computed on the
core scope on which credibility scores are produced. The per-m$^2$/yr
disclosure rate (Section~\ref{sec:gap}) is reported with its
sensitivity to including these building-services supply-chain firms
noted as a scope-robustness check.

\subsection{The BeDA instrument}\label{sec:instrument}
BeDA is a four-phase pipeline (Fig.~\ref{fig:pipeline}). The design
principle is reliability-before-use: no credibility score is interpreted,
and no downstream gap statistic is reported, until the score has been shown
to be model-robust and confound-robust. Each phase is described in turn.

\textbf{P1 --- Credibility scoring.} Every report is rendered to a
multimodal context (page text plus page imagery) and scored by
\texttt{gemini-3.1-pro-preview} against the fixed SSSR rubric
(Table~\ref{tab:rubric}). The model returns the topic-breadth score
(A), the standards-compliance spine (B), and three multimodal
sub-scores ($C_1,C_2,C_4$). The spine is the credibility score on which the audit
rests. The topic-breadth score and the multimodal composite
($C=C_1+C_2+C_4$) are carried through P2 so that their reliability
can be contrasted with the spine's. The primary-model scores are reused
from the SSSR corpus without re-scoring, so P1 introduces no new model
variance.

\textbf{P2 --- Reliability battery.} A reliability battery is a set of tests that check whether the score is
reproducible: does it change when the model producing it changes? We run
four tests.
(i)~\emph{Cross-model}: every shared built-environment report is
independently re-scored by \texttt{gemini-3-pro-preview} and the
spine concordance computed (Pearson and Spearman).
(ii)~\emph{Cross-family}: the same reports are re-scored by two Gemma-4
models and two Gemini-flash models (Table~\ref{tab:models}). Agreement
that survives a change of model \emph{family} (different pre-training,
albeit the same vendor) is the strongest evidence that $B$ measures the
underlying construct. Gemma models are run text-only, which additionally
yields a conservative lower bound on text-channel reliability.
(iii)~\emph{Leakage}: to test whether the score reflects the document
content or pre-trained knowledge of the firm
\citep{carlini2021extracting,carlini2023quantifying}, a 30-report
subsample is
re-scored with all firm identifiers removed and the original-versus-anonymised
$B$-correlation measured.
(iv)~\emph{Cross-vendor}: the frozen 100-report audit subset is
re-scored by a non-Google vision model (\texttt{qwen3.6-27b}) from
byte-identical page renders with the byte-identical prompt and rubric,
and on the text channel by three non-Google models
(\texttt{deepseek-chat}, \texttt{minimax-m2.7}, \texttt{qwen3.6-27b}),
three runs per arm. The primary comparison, its falsification
threshold ($r \geq 0.70$ on the spine), a length-confound test in every
arm, a degenerate-output flag and two directional predictions were
registered to disk before the comparator arm was scored (the final
registration file is timestamped 72\,s ahead of the first request).

\textbf{P3 --- Confound-controlled validity.} A \emph{confound} is a
third variable that can inflate an apparent association: here, chiefly
how much a firm discloses (operationalised as log report length, the
natural logarithm of one plus the word count of the report's extracted
text),
together with its listing region (US/non-US), sector division and
reporting year. We test whether $B$ agrees with an independent ESG
rater and whether it is associated with recorded environmental
incidents, using regressions that hold those variables constant, with
standard errors clustered by firm. Much of the reviewed literature
reports the raw correlation as validity
(Section~\ref{sec:baselines}); we do not. A separately produced
two-coder human gold set ($n=100$, SJTU, six sectors, pre/post-2018)
provides a head-to-head measurement comparison reported in
\ref{app:goldset}.

\textbf{P4 --- Disclosure-fitness application.} The validated
instrument is then applied to a decision-relevant question: what
fraction of built-environment disclosure can support science-based
stranding assessment? The design has two layers. An LLM extractor first
reads each full report PDF and returns the disclosed metric fields
(intensity value and unit, energy intensity, absolute Scope~1+2, floor
area, reporting year). Whole-PDF long-context reading is a stated
design point rather than a default: on 9{,}369 Chinese annual reports,
DeepGreen reports an ablation in which ``Retrieval-Augmented Generation
(RAG) reduces hallucinations, as compared to simply lengthening the
input window'' \citep{deepgreen2026}. We keep the whole document
because the target fields sit in tables and charts whose page images a
text retriever discards, and we bound the hallucination that retrieval
is meant to suppress by literal grounding instead ($97.3\%$,
Section~\ref{sec:gap}); a three-arm ablation (whole-PDF multimodal,
whole-document text, retrieval over the same text) on the $n{=}80$
second-extractor subset is offered at revision. A strict extract-only instruction returns null
for anything not explicitly disclosed --- the absence convention that
evidence-grounded emissions-extraction benchmarks also impose on their
human annotators (``Annotators do not infer undisclosed categories;
missing categories are recorded as null'' \citep{scope3trace2026}).
The convention carries a limitation that Forster et al.\ \citep{forster2026}
state for their European extraction and that applies equally here:
``absence of an extracted value does not necessarily represent the
absence of reporting.'' Because the cascade's headline is a rate of
absence, we treat recall on the null stratum as a validity question in
its own right, answered by a blinded human audit of
non-disclosing reports (Section~\ref{sec:limitations}) rather than by
agreement between two extractors, which cannot see what both miss. A deterministic
rule-based classifier (regular expressions plus a unit parser, with a
physical-plausibility cap) then assigns each report to one of seven
disclosure-fitness classes: operational carbon per m$^2$/yr
(CRREM-fit), energy-use intensity per m$^2$/yr (EPBD-fit), per-unit or
per-home (embodied), revenue/output-normalised, absolute Scope~1+2
without intensity, no usable intensity, and implausible or extraction
error. This yields the per-m$^2$ reporting gap of
Section~\ref{sec:gap}; implementation details are in
\ref{app:implementation}.

\textbf{P5 --- City bridge (cross-arm reliability).} To connect the
entity-level cascade to the jurisdictions that regulate buildings, a
second extraction pass reads each real-estate report for the cities,
metropolitan areas and states in which the owner names holdings and
matches them against enacted building performance standards. The unit
is the owner; exposure is the union of matches over all of an owner's
reports; status is read from the latest FY$\geq$2020 report. The pass is
run in four arms: the primary (\texttt{gemini-3.1-pro-preview}, whole
PDF); a later same-vendor model on the same whole-PDF stimulus
(\texttt{gemini-3.7-flash}); that model on extracted text only; and a
non-Google model on the stimulus-matched text (DeepSeek~V4~Flash).
Three cross-arm predictions were registered before the arms were
scored --- three-way verdict agreement of at least 90\% for the
same-stimulus model pair, 85\% for the cross-vendor text pair and 90\%
for the modality pair --- and none was met (80.9\%, 74.2\% and 87.2\%
respectively). The disagreements concentrate on whether a report names
any market at all; the binary exposure verdict the bridge rests on
agrees at 93--96\% report-level across arms, and a 30-report retest of
the primary agrees at 93.3\% three-way. The unmet predictions are
reported as such, the primary arm's numbers are not moved, and the
owner set exposed under every arm (22 owners) is reported alongside
the primary set as the reproducible core. As with P4, cross-arm
agreement is reproducibility, not validity: validity rests on literal
grounding of the extracted market names in the source PDF (96.6\% of
2{,}675 decidable names) and on a blinded 40-report hand check of the
exposure and readability verdicts. That check failed on the final
codes on all four of its registered criteria (Section~\ref{sec:limitations}), so under
the rule registered before it ran no city-bridge number is reported as
a finding; the extraction is described here as method, and the
exposed-owner counts are withheld.

\begin{table}[!htbp]\centering
\caption{Model roster: the Gemini family \citep{gemini2023,gemini2025}
and the Gemma family \citep{gemma2025,gemma2026}. The spine is reused (already scored);
reliability is tested by independent re-scoring across families.}
\label{tab:models}
\footnotesize\setlength{\tabcolsep}{4pt}
\begin{tabular}{@{}p{0.36\linewidth}p{0.17\linewidth}p{0.43\linewidth}@{}}
\toprule
Model & Family & Role\\
\midrule
gemini-3.1-pro-preview            & Gemini & Primary (spine)\\
gemini-3-pro-preview              & Gemini & Cross-model validator (within-Pro, multimodal)\\
gemini-3-pro-preview (text-only)  & Gemini & Modality-control validator (same Pro, vision off)\\
gemini-3-flash-preview            & Gemini & Cross-tier validator (Pro $\to$ Flash, 3-series)\\
gemini-3.5-flash                  & Gemini & Cross-tier validator (Pro $\to$ Flash, 3.5-series)\\
gemma-4-31b-it                    & Gemma  & Cross-family validator (text-only)\\
gemma-4-26b-a4b-it                & Gemma  & Cross-family validator (text-only)\\
qwen3.6-27b (vision)              & Qwen   & Cross-vendor validator (pre-registered, 100-report subset)\\
deepseek-chat, minimax-m2.7, qwen3.6-27b (text) & DeepSeek / MiniMax / Qwen & Cross-vendor text-channel arms (sensitivity)\\
\bottomrule
\end{tabular}
\end{table}

\subsection{Implementation}\label{sec:implementation}
The built-environment flag is assigned by a crosswalk that maps each firm to
a US SEC Standard Industrial Classification code where available and to a
global ticker-based industry lookup otherwise. The crosswalk is deliberately \emph{three-state} (built-environment, not
built-environment, or unresolved) because silently coercing the
unresolved tail to ``not built-environment'' would bias the gap
statistic downward. Unresolved firms (485 of 2{,}246, 21.6\% of the
panel; 66.4\% of them US-listed, close to the panel's 71.5\%) are
carried explicitly and handled by rate-extrapolation: the
built-environment rate observed among resolved firms is applied to the
unresolved share. The assumption is that unresolved firms contain
built-environment firms at the same rate as resolved firms, and the
unresolved tail's listing composition being close to the panel's is
what that extrapolation needs; at the 7.2\% base rate it implies
roughly 35 missed built-environment firms, which is why ``census'' is
used throughout as census \emph{of the resolved frame}. This gives a floor of 116 built-environment firms
(treating all unresolved as non-built-environment) against a
rate-extrapolated estimate of 148. Of the two pre-registered
feasibility conditions, the first was met (the built-environment
stratum reaches a usable size: $n=116$ on the core scope, 127 tagged
on the expanded scope, 119 of them with collected reports) and the
second \emph{failed}: the pre-registered gate required a
CRREM-constructible share of at least 0.35 within the panel stratum,
the observed share was 0.20, and the documented pivot was to assemble
the separate real-estate corpus of Section~\ref{sec:readiness}, whose
218 constructible reports power the stranding screen. We state this
because a failed gate with a documented pivot is a different
epistemic object from a passed gate. All stages (crosswalk, scoring, reliability
battery, confound-controlled regressions, and gap classifier) are
scripted end to end and reproduce from the analysis code described
under Data availability, with all model
versions pinned (Table~\ref{tab:models}). The related work in Section~2
was identified by citation snowballing from seed references and screened
to high-impact venues; every cited source was verified against
Crossref.

\subsection{Baselines, metrics, and validation design}
\label{sec:baselines}
\textbf{Baselines.} Every comparator is either evaluated with the
identical confound-controlled validity tests or excluded for a stated
identification reason (Table~\ref{tab:baselines}; results in
\ref{app:validity}). The Loughran--McDonald financial-sentiment
lexicon is the classical text-only benchmark. ClimateBERT
\citep{bingler2022cheap,bingler2024cheaptalk} is the published
deep-learning comparator,
evaluated zero-shot on the full panel with its authors' own cheap-talk
construct: a detector--specificity--commitment cascade yielding a
specificity share and a cheap-talk (commitment-without-specificity) share
per report. A dimension ablation contrasts the standards-compliance spine alone with
a spine-plus-multimodal composite ($B{+}C_1{+}C_2{+}C_4$) to test
whether the multimodal dimensions add validity. The input-channel
ablation is the text-only Pro arm of
Section~\ref{sec:reliability}. A fine-tuned LSTM/BERT is the one
comparator that is not estimable here: supervised fine-tuning needs
panel-scale credibility labels, which do not exist. The only human
labels (the $n{=}100$ gold set) are range-restricted
(\ref{app:goldset}), so fine-tuning to them would validate against
a compressed target, and distilling from $B$'s own scores would be
circular. Zero-shot ClimateBERT is therefore the strongest estimable
deep-learning comparator.

\textbf{Metrics.} Concordance uses Pearson $r$ (linear agreement) and
Spearman $\rho$ (rank agreement). For the LLM-versus-human and
inter-coder comparisons we add quadratic-weighted Cohen's $\kappa$
\citep{cohen1960,cohen1968} on
score quintiles, which penalises large disagreements more heavily;
agreement magnitudes are interpreted against the Landis--Koch
benchmarks \citep{landiskoch1977}. The convergent criterion is the
LSEG ESG Score (the pillar-weighted score, not the
controversies-combined ESGC), joined to reports by ISIN and reporting
year. The incident criterion is the firm-level count of environmental
risk incidents recorded by RepRisk for the panel firms, as provided in
the panel's incident extract. Convergent validity is reported at three
levels of control: raw Pearson $r$, partial $r$ after volume
adjustment, and a fully controlled regression coefficient with
firm-clustered bootstrap 95\% confidence intervals, so the reader can
see attenuation explicitly. The incident relationship is estimated at
the firm level (incident counts against the firm's mean spine score)
with a negative-binomial regression (appropriate for over-dispersed
counts) and reported as an incidence-rate ratio (IRR). An IRR above one
means a higher credibility score is associated with proportionally more
recorded incidents, net of controls. The cascade rates are presented
as exact shares of their stated frames with firm-clustered
bias-corrected-and-accelerated (BCa) bootstrap intervals (20{,}000
draws, seed recorded, bounds to the nearest whole point). Because
Frame A is a census, its intervals are \emph{model-based} ---
uncertainty over a superpopulation of firms, not sampling error in the
census count --- and are labelled so wherever quoted. Every stage
definition, the pre-committed diversified-portfolio rule, and the
data-corrections policy are fixed in a written measurement
specification released with the code, and every number in this paper
is regenerated from the raw extraction outputs by a claims registry
that fails loudly on drift.

\textbf{Validation design.} All inferential analyses target the
built-environment lead stratum; the full panel is retained as the
comparison frame that carries inferential weight. Stratum-versus-panel differences are tested by an
interaction term on the pooled sample, which has more statistical power
than comparing two small-sample correlations. Sample sizes were set by
design targets and pre-specified feasibility gates
(Section~\ref{sec:implementation}) rather than a formal power
calculation: the $n{=}200$ stratified sample was sized for precision,
yielding a Wilson 95\% interval of roughly $\pm$5--6 percentage points
at the anticipated rate near 20\%. Sample sizes are stated per analysis
and small-$n$ caveats are confined to Limitations.

\begin{table}[!htbp]\centering
\caption{Baseline roles. Every comparator is either evaluated on the same
confound-controlled validity battery (\ref{app:validity}) or excluded
for a stated identification reason (Section~\ref{sec:baselines}).}
\label{tab:baselines}
\small
\begin{tabular}{lll}
\toprule
Baseline & Type & Status\\
\midrule
Loughran--McDonald lexicon & classical text floor & evaluated (\ref{app:validity})\\
ClimateBERT (zero-shot cascade) & published deep-learning comparator & evaluated (\ref{app:validity})\\
Spine-only vs spine + multimodal & dimension ablation & evaluated (\ref{app:validity})\\
Fine-tuned LSTM/BERT & supervised deep learning & not estimable (no valid labels)\\
\bottomrule
\end{tabular}
\normalsize
\end{table}

\section{Results}
The primary results are the panel-scale disclosure-fitness gap and the
stranding-readiness it implies (Sections~\ref{sec:gap}
and~\ref{sec:readiness}). The reliability and validity evidence
presented first is the \emph{credential} for those measurements: it
shows that automated multimodal-LLM auditing of long, mixed-format
disclosures is trustworthy enough to act on. The same credential covers
both the credibility score and the disclosure-fitness classification on
which the headline gap rests. The leakage probe, LLM-versus-human
measurement study, baseline validity comparison, and CRREM construction
details are in the appendices.

\subsection{Reliability of the enabling instrument}\label{sec:reliability}
\begin{figure}[!htbp]\centering
\includegraphics[width=.62\linewidth]{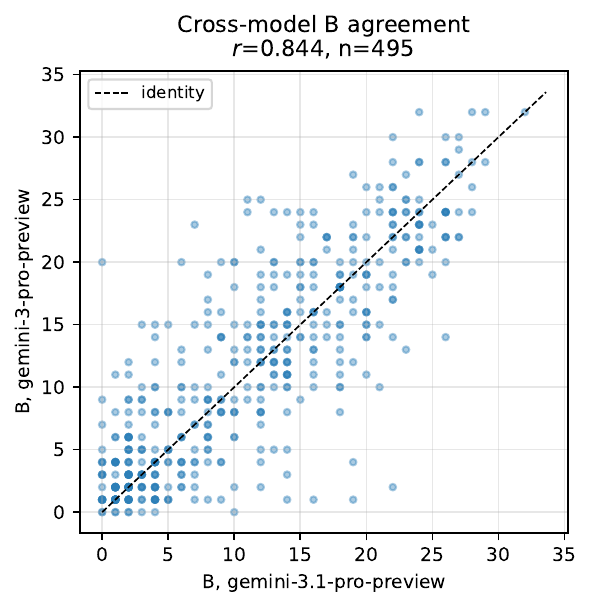}
\caption{Cross-model agreement of the credibility spine between
\texttt{gemini-3.1-pro-preview} (primary) and \texttt{gemini-3-pro-preview}
(within-Pro validator) on the full common-scored set ($n=495$,
$r=0.844$); the built-environment subset gives $r=0.845$ ($n=407$).}
\label{fig:xmodel}
\end{figure}

The standards-compliance spine is cross-model stable: Pearson
$r=0.845$ (Fisher 95\% CI $[0.81, 0.87]$) on the $n=407$
built-environment reports scored by both
\texttt{gemini-3.1-pro-preview} and \texttt{gemini-3-pro-preview} (full
common-scored set: $495$ reports, $r=0.844$; Fig.~\ref{fig:xmodel}).

\begin{figure}[!htbp]\centering
\includegraphics[width=.7\linewidth]{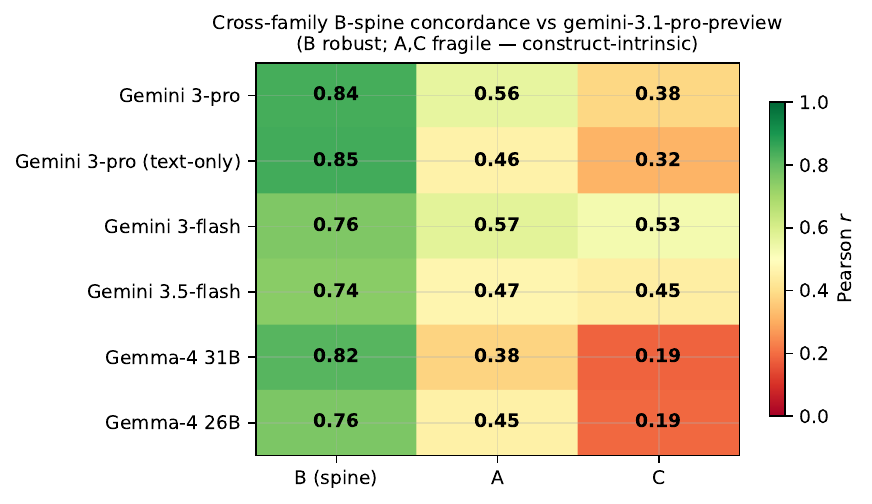}
\caption{Cross-family concordance vs the primary
\texttt{gemini-3.1-pro-preview}. The standards-compliance spine stays
high across the Gemma family and Flash-tier Gemini variants;
topic-breadth and the multimodal scores collapse. Short labels identify
the specific cross models; full identifiers in
Table~\ref{tab:models}.}
\label{fig:xfam}
\end{figure}

\begin{figure}[!htbp]\centering
\includegraphics[width=.78\linewidth]{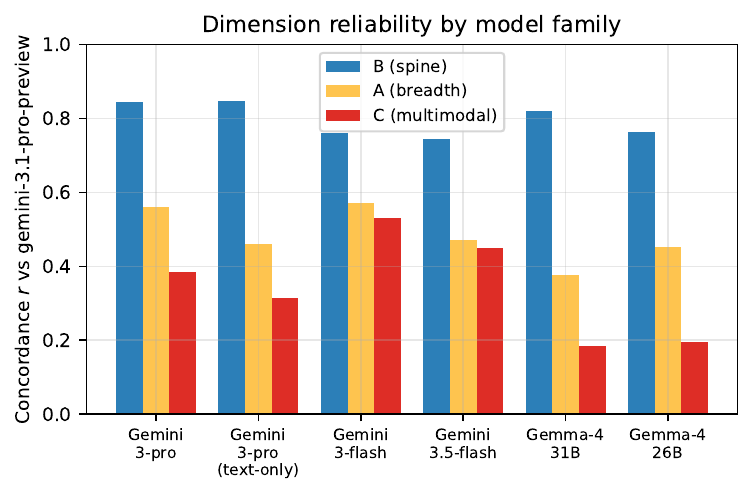}
\caption{SSSR dimension reliability by model. Only the
standards-compliance spine retains high concordance across model
families; the fall in topic-breadth and multimodal scores against
text-only validators is largely attributable to the absence of a shared
visual channel, as isolated by the text-only Pro arm
(Section~\ref{sec:reliability}).}
\label{fig:dimbars}
\end{figure}

The result holds \emph{cross-family} (Fig.~\ref{fig:xfam},
Table~\ref{tab:xfam}). On a shared 100-report audit subset the
standards-compliance spine stays at $r\approx0.74$--$0.82$ against the
Gemma-4 family and the Flash-tier Gemini variants
(\texttt{gemini-3-flash-preview} and \texttt{gemini-3.5-flash}, the
latter from a subsequent release cycle). The lowest of these
coefficients ($r{=}0.744$ at $n{=}99$; Table~\ref{tab:xfam}) carries a
Fisher 95\% CI of $[0.64, 0.82]$. The topic-breadth and
multimodal-evidence scores collapse to $0.19$--$0.57$
(Fig.~\ref{fig:dimbars}). This cross-family agreement is consistent with
the spine measuring a shared underlying construct, but every validator
above is a Google model, so it cannot by itself exclude a vendor-level
artefact. The pre-registered cross-vendor arm (P2(iv)) closes that gap
within stated limits. On the same frozen 100-report audit subset, a
non-Google vision model (\texttt{qwen3.6-27b}) reproduces the primary
spine at Pearson $r{=}{+}0.725$ and Spearman $\rho{=}{+}0.812$
($n{=}100$, median of three runs). The Pearson value \emph{narrowly}
clears the pre-registered $0.70$ threshold (margin $+0.025$; the
first-run-only estimate of $+0.706$ is borderline, and the weakest
text-channel arm, \texttt{minimax-m2.7} at $r{=}{+}0.670$ against the
text-only primary, would not clear it). Four qualifications are
mandatory. First, the subset is an extreme-groups draw (49
bottom-quintile, 49 top-quintile and 2 mid-range reports), so every
agreement statistic on it is range-inflated relative to a random
corpus draw and is comparable only within these 100 reports; the
within-vendor $r{=}0.845$ at $n{=}407$ remains the panel-scale
estimate. Second, agreement bounds vendor-dependence only, not
validity: on this benchmark subset the spine is \emph{negatively}
rank-correlated with expert-consensus scores (Spearman
$\rho{=}{-}0.360$, $n{=}100$), the same sign as the LLM-versus-human
divergence of \ref{app:goldset}, and no model-to-model concordance
rebuts that. Third, the association between the spine and report
length, which motivates the log-length control in every validity test
(Section~\ref{sec:baselines}), is present in all eight cross-vendor
arms on this sample ($\rho(B,\mathrm{pages}){=}{+}0.478$ to
${+}0.720$; corpus-level ${+}0.545$ at $n{=}6{,}993$): it is a
property of LLM evaluators under this stimulus, not a Gemini artefact.
Fourth, in this design page capture rather than model identity is the
dominant driver of disagreement on the multimodal sub-scores: the same
Flash model scored from a different page draw keeps the spine at
$\rho{=}{+}0.886$ while $C_1$, $C_2$ and $C_4$ fall to
$+0.348$--$+0.557$ ($n{=}99$), which independently corroborates the
page-sampling sensitivity of \ref{app:leak} and the deterministic
page-seed design. One further fact belongs here as a live instance of
the deprecation argument below: \texttt{gemini-3-pro-preview}, the
within-Pro validator of Fig.~\ref{fig:xmodel} and the second extractor
behind the census $\kappa{=}0.95$, was retired by the vendor
mid-project (HTTP 404, verified 9 August 2026); its arms are archived
and reproducible from the released scores but can no longer be re-run
at the vendor. Taken with the cross-family battery, the cross-vendor
replication completes the reliability evidence. It
is the strongest reliability evidence available short of an external
ground truth: among the three SSSR dimensions, only the
standards-compliance spine retains high concordance across model
families.

For regulators and portfolio owners this is the key practical property.
Either can switch to a newer scoring model without
the credibility ranking shifting materially. Vendors deprecate and
replace LLMs on a months-long cycle, as illustrated here by a Flash
variant from a later release. Stability across that turnover is a
necessary (though not sufficient) condition for use in a Local Law~97 or
EPBD triage workflow.

We separate the fall in topic-breadth and multimodal concordance from a
modality artefact using a text-only Pro arm
(\texttt{gemini-3-pro-preview} re-scored with the visual channel
disabled on the same 100-report subset). Against the multimodal primary,
the text-only Pro arm yields a spine concordance of $0.847$
(essentially identical to the multimodal pair's 0.844), a topic-breadth
concordance of $0.460$ (down from 0.562), and a multimodal-composite
concordance of $0.316$ (down from 0.384). The comparison supports two
readings. First, the spine is robust to both family \emph{and}
modality: the construct claim survives. Second, the topic-breadth
decline is \emph{largely modality-driven}: removing vision drops its
concordance to a level comparable to the text-only Gemma validators.
The multimodal score is mixed: the text-only Pro arm loses some
concordance (consistent with the visual nature of C1/C2/C4) yet stays
above the Gemma family, indicating a small additional construct
component beyond modality. BeDA accordingly confines every substantive
claim to the spine. The multimodal input channel is retained for a
practical reason: these are long PDFs whose quantitative content sits in
tables and figures, exactly the tabular intensity disclosures that P4
extracts. The multimodal scores serve as the reliability contrast that
shows what does \emph{not} survive.

Both Flash-tier validators show a systematic upward bias on the spine
(mean absolute error ${\approx}6$ on the $0$--$32$ scale, against
${\approx}2$ for Pro). We therefore retain Pro as the cross-model audit
head; the Flash variants enter only as cross-tier reliability evidence.
The reproducibility of this pattern across two distinct Flash releases
is worth flagging for any future deployment that contemplates a
Flash-only pipeline as a cost optimisation.

\subsection{Confound-controlled validity}\label{sec:validity}
\begin{figure}[!htbp]\centering
\includegraphics[width=.58\linewidth]{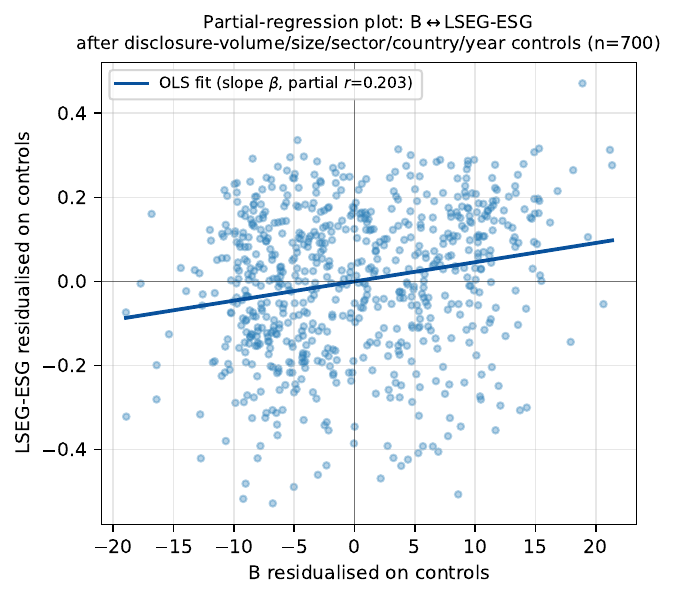}
\caption{Partial-regression (added-variable) plot of the credibility spine
B against LSEG-ESG after partialling out disclosure volume (log report
length), listing region, sector division and reporting year. The non-zero slope is the confound-controlled
association; point estimates, CIs, the built-env$\times$panel interaction
and the RepRisk IRR are in Table~\ref{tab:valid}.}
\label{fig:validity}
\end{figure}

As expected, raw bivariate associations attenuate once disclosure
volume (log report length), listing region, sector division and
reporting year are controlled (Fig.~\ref{fig:validity},
Table~\ref{tab:valid}). Convergence with the independent ESG rater
falls from a raw $r=0.236$ to a volume-adjusted partial $r=0.203$ and a
small but significant fully controlled coefficient
($p\approx6\times10^{-5}$). The built-environment$\times$panel
interaction is significant ($p\approx2\times10^{-6}$, pooled $n=700$),
indicating a detectable stratum-versus-panel difference. The
association with recorded environmental incidents survives controls
(negative-binomial IRR${=}1.05$, $p\approx6\times10^{-9}$,
$n=1{,}052$). We therefore
report convergence as modest. The head-to-head against a
two-coder human gold set is reported in \ref{app:goldset}.

In practice, the credibility score is defensible as a \emph{screening
tool} that helps a regulator rank thousands of disclosures for closer
review, but not as a standalone verdict on any single firm. Because disclosure volume is operationalised as log report
length, this surviving convergence is not a verbosity artefact: the
external criteria are recovered net of how much a firm writes.

Neither text-only baseline recovers the spine construct: the
Loughran--McDonald lexicon reaches $r{=}0.21$ at best, and a zero-shot
ClimateBERT cascade only $r{=}{-}0.09$ panel-wide. On the identical
confound-controlled tests, only the spine-based scores converge across
all three external criteria simultaneously
(Table~\ref{tab:baseline_validity}, \ref{app:validity}). A
dimension ablation shows that adding the multimodal sub-scores weakens
every point estimate (LSEG partial $r$ of $+0.203$ for the spine alone
versus $+0.109$ for the spine-plus-multimodal composite, overlapping
CIs), which is why every substantive claim is confined to the spine.

The incident signal, by contrast, survives every control. Net of
disclosure volume, listing region and sector, a higher compliance score
is associated with proportionally \emph{more} recorded environmental
incidents (IRR${=}1.05$). This direction is consistent with the
greenwashing pattern reported in the firm-level LLM literature, though
it is not established here: RepRisk incident counts also scale with
firm prominence and media coverage, which the control set only
partially proxies. The association remains the practically relevant
signal for an owner screening transition risk, and the kind of
confound-controlled test that the greenwashing-detection literature
(Section~2.2) omits.

\subsubsection*{External-criterion convergent validity (built-env-specific)}
Beyond the generic external criteria above (LSEG-ESG, RepRisk
environmental incidents), we test \emph{built-environment-specific}
convergent validity against SBTi near-term target validation status,
a binary, public, weekly-updated
criterion with explicit sectoral classifications for real-estate,
construction, building-products and electrical-equipment firms
(\texttt{sciencebasedtargets.org}, accessed 2026-05-19, i.e.\ under
Version~1.1 of the SBTi Buildings Sector Criteria; Version~1.2,
effective 20 July 2026, records only clarifications of
emissions-inventory coverage and the upfront embodied-emissions
boundary plus administrative changes, leaving the per-m$^2$ intensity
denominator and pathway-selection rules as cited above
\citep{sbti2026buildings}). On the
expanded built-environment panel ($n{=}127$ tagged firms, 119 with a
collected report; Section~3), $34$ firms are present in the SBTi
public dashboard (matched by ISIN
where available, otherwise by company-name token-Jaccard
$\geq 0.67$). Of these, $27$ have a validated near-term target. On the
$n{=}119$ firms with at least one credibility score in our scored corpus,
\textbf{the point-biserial correlation (a Pearson $r$ between the mean
$B$ score and SBTi-validation status) is $0.357$}
($p{\approx}7\times10^{-5}$; $+0.360$ on the firm-mean join of the
baseline matrix, Table~\ref{tab:baseline_validity}; Mann--Whitney
$p{\approx}3\times10^{-4}$). The mean B is $13.24$ for SBTi-validated
firms ($n{=}24$) versus $7.53$ for those not validated ($n{=}95$), a
$76\%$ relative gap
($5.71$ points on the $0$--$32$ scale). The effect is medium-magnitude
by conventional benchmarks \citep{cohen1988} ($r{\approx}0.35$) and
converges in the same direction as the LSEG-ESG
and RepRisk tests: an entirely independent, built-environment-relevant
signal recovers the $B$ ordering. This test is a raw point-biserial, not
confound-controlled like the LSEG/RepRisk estimates. SBTi validation is
moreover an imperfect criterion in its own right: reported progress
against science-based targets can be inflated, for example by
renewable-energy certificates \citep{bjorn2022recs}. This is a further
reason it serves here as corroboration only. CDP Climate
and GRESB Real Estate Assessment scores require subscription access and
are a natural extension when available.

\subsubsection*{The human gold set: an LLM-versus-human measurement note}
A two-coder human gold set (SJTU; 100 reports) is internally reliable
(B $w\kappa{=}0.857$), yet its agreement with the LLM B-score is
\emph{negative} ($r{\approx}{-}0.24$--${-}0.29$). We show in
\ref{app:goldset} that this is a leniency phenomenon: the
human coders compress to the upper third of the scale while the spine
spans it. The pattern replicates in an independent LLM pipeline
($r{\approx}{-}0.38$) and in the Loughran--McDonald lexicon (lexicon vs
LLM $r{=}{+}0.44$; vs human $r{=}{-}0.10$; \ref{app:goldset}).
Three independent automated methods thus diverge from the human scores
in the same direction. This is most consistent with the divergence
arising in human coding of long structured disclosures. Construct
validity is therefore anchored on the three external criteria, not on
the human pass. We cannot fully exclude a shared sensitivity to
compliance-correlated surface features, although the convergence
survives a log-report-length control, which rules out simple verbosity
(Section~\ref{sec:limitations}).

\subsection{Robustness}
We also test for \emph{leakage}, the risk that the score reflects the
firm's identity rather than the content of its report. The spine is
robust on both input channels (text-channel $r=0.886$; vision-channel
$r=0.826$ under cover-page omission), with only a small page-sampling
sensitivity on the visual channel (${\approx}{+}2$ B). Details and the
figure are in \ref{app:leak}.

\subsection{The constructibility cascade (primary result)}\label{sec:gap}
The census covers all 502 collected reports of the 119
built-environment firms with at least one report among the 127 tagged
on the expanded scope (Section~3; FY\,2003--2023, 494 cleanly
extracted); the
real-estate corpus is described in Section~\ref{sec:readiness}. Three
documented data corrections (unit typos confirmed by the rows' own
evidence notes; one chart misread settled by a PDF word-coordinate
audit) are applied at load from a released corrections file; all
reliability statistics are computed on uncorrected extractor output.
The classifier reliably extracts the disclosure-fitness
disaggregation of Section~\ref{sec:instrument} (Cohen's
$\kappa{=}0.95$, bootstrap 95\% CI [0.87, 1.00], on a 50-report subset
re-scored by an independent extractor; and $97.3\%$ of all extracted
intensities appear verbatim in their source PDF).

\begin{figure}[!htbp]\centering
\includegraphics[width=.98\linewidth]{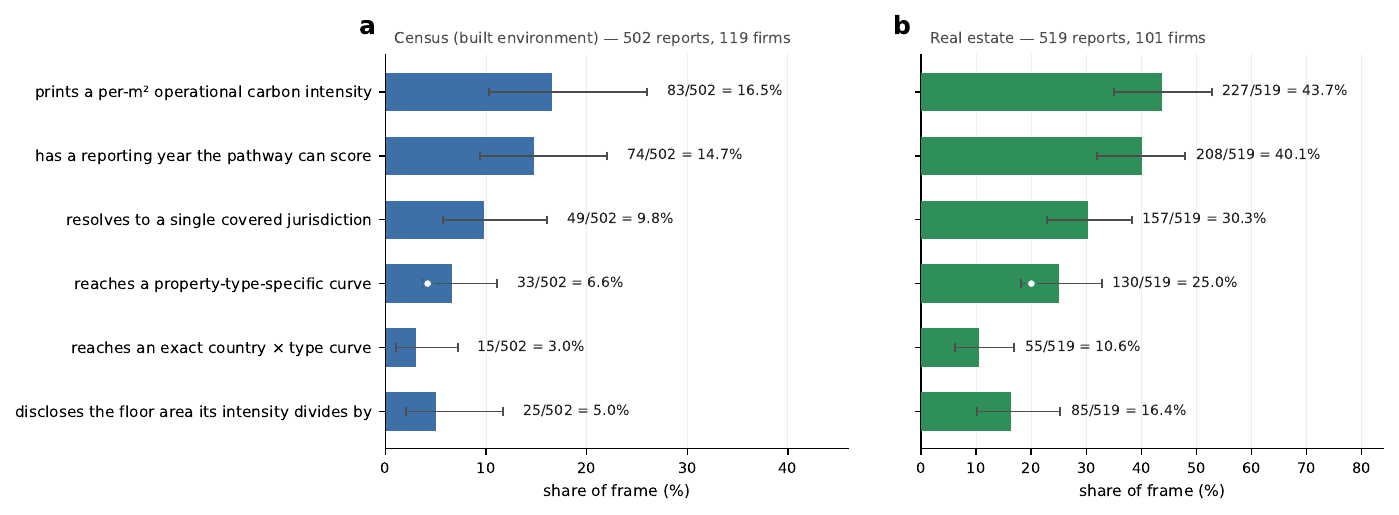}
\caption{The constructibility cascade on the two frames (printed-only
basis; firm-clustered BCa 95\% intervals, 20{,}000 draws, model-based
for the census). Open circles mark the strict single-asset-class
sensitivity at the property-type stage (the lenient rule is primary;
the gap between them is 12 census and 26 real-estate multi-asset
reports). The census's floor-area stage (5.0\%) sits \emph{below} its
property-type stage (6.6\%).}
\label{fig:cascade}
\end{figure}

\textbf{16.5\% of census firm-reports (83/502; firm-clustered BCa 95\%
[10\%, 26\%], model-based) print an operational carbon intensity per
m$^2$/yr} --- printing by the firm, not derivability by an analyst: a
further 7 reports (1.4\%) permit a scope-1+2\,$\div$\,floor-area
derivation and are counted separately. In the real-estate corpus the
printed rate is 43.7\% (227/519). The cascade
(Fig.~\ref{fig:cascade}) then runs the pathway frameworks' own
admission gates over the printed disclosures. In the census: 14.7\%
survive the pathway window; 9.8\% resolve to a single covered
jurisdiction\footnote{``Jurisdiction'' is used in CRREM's own sense:
``the geographic unit for which CRREM publishes Pathways. This may be a
country or a sub-national jurisdiction (e.g.\ sub-country climate zones
as used in Australia, or blended climate and electricity/grid regions
as used in the United States)'' \citep{crrem2026euiexplainer}. For the
United States these are 57 sub-national regions, one per ``unique
combination of eGRID region and CBECS climate zone'', which the CRREM
tool retrieves from an asset's ZIP code \citep[p.~22]{lbnl2024crrem}.
The single-jurisdiction gate asks only whether a report resolves to one
country; no report in either corpus discloses a US sub-national region,
which is why the US carbon verdict is bounded rather than identified
(Section~\ref{sec:grid}).}; 6.6\% obtain a property-type-specific curve (4.2\% under
the strict single-asset-class rule --- the lenient rule, under which a
blended multi-asset ratio still counts, is primary because ambiguity
is resolved \emph{against} our thesis); 3.0\% reach an exact
country$\times$property-type curve; and \textbf{5.0\% disclose the
floor area their own intensity is divided by}. The real-estate chain
runs 43.7\% $\to$ 40.1\% $\to$ 30.3\% $\to$ 25.0\% (strict 20.0\%)
$\to$ 10.6\%, with 16.4\% disclosing the denominator. One inversion
does the argumentative work of a paragraph: in the census,
\emph{fewer} reports disclose their own floor area (5.0\%) than can be
matched to a property-type pathway (6.6\%) --- benchmarkable is not
the same as verifiable, because a match to a curve does not let anyone
check the intensity's own denominator. At the firm level, 26 of 119
census firms (61 of 101 real-estate firms) print the intensity in at
least one report; among firms with three or more reporting years, 6 of
64 (17 of 86) print it in every year.

\textbf{Why printed intensities still fail the pathway gates.} Of the
census reports that print an intensity but obtain no
property-type curve, 82\% fail on the disclosure side --- the report
blends jurisdictions or asset classes into one entity-level ratio, or
names no property type --- and 18\% fail on the library side (a named
single jurisdiction or type that CRREM V2.04 does not cover). The
real-estate decomposition is 83\% versus 17\%. The obstacle is the
reporting unit, not missing coverage: entity-level disclosure reports
one blended ratio where the benchmark is defined for one asset type in
one jurisdiction, and a mandate to publish one more number does not
disaggregate a portfolio.

\textbf{The regional gradient is a standards gradient, not a
geography gradient.} The printed rate is 65.4\% (51/78) for
London-listed and 71.0\% (49/69) for continental-European real-estate
reports against 35.2\% (102/290) for US listings, consistent with the
EPRA sBPR industry code, whose current edition requires data coverage
to be reported for each asset-level performance measure
\citep{epra2024sbpr}, and with the per-m$^2$ basis of the EPBD recast
(though the ESRS~E1 mandatory intensity metric is revenue-based
\citep{eu2023esrs}, and its 2025 amendment codifies an
internal-estimates fallback for collateral energy data
\citep{efrag2025e1}). Conditioning bears this out: in a firm-clustered
linear probability model on the real-estate corpus (FY$\geq$2016,
$n{=}459$, US-only reference, 101 firm clusters), the UK/EU and
continental coefficients ($+0.29$, $p{=}0.040$; $+0.37$,
$p{<}10^{-4}$) collapse to non-significance ($-0.01$; $+0.15$,
$p{=}0.23$) once two standards terms are added, while referencing the
GRESB benchmark carries $+0.30$ ($p{\approx}10^{-6}$, the one
coefficient that survives Holm adjustment across every test in this
paper) and the EPRA term ($+0.35$, $p{=}0.003$) is close to
tautological, since the sBPR code defines the per-m$^2$ metric. The
lever is the benchmark a firm reports into, not the jurisdiction it
lists in. Two comparisons we explicitly decline to make claims from:
the census rate for FY2015--19 versus FY2020--23 (20.4\% vs 18.5\%,
Fisher $p{=}0.70$, power 0.57 against a rise to 30\%) cannot
distinguish flat from rising, and the firm-clustered trend
($+1.0$\,pp/yr, raw $p{=}0.040$) does not survive Holm adjustment;
both are reported descriptively. Sensitivities: dropping the 31
report-years the two frames share moves the census printed rate to
14.9\% and the real-estate rate to 43.4\%; the recency-selected
200-report census arm prints at 20.1\% against 14.7\% for the
remainder (Fisher $p{=}0.09$, descriptive --- 61 firms appear in both
arms).

For cities and portfolio owners, Local Law~97 and the EPBD recast
presuppose that a building's carbon intensity can be compared against a
declining science-based pathway. Yet five in six census firm
disclosures do not print intensity on the floor-area basis that
comparison requires, and of those that do, most cannot be carried to a
named curve. Enforcement is therefore bottlenecked at the disclosure
source; better downstream modelling cannot remove the bottleneck.
BeDA's contribution to the regulator is two-fold: it screens
\emph{which} disclosures are credible (the reliability-validated
spine) and it localises \emph{where the decision-useful-data gap is}.
A mandate to close that gap --- intensity disaggregated by asset type
and jurisdiction, reported with the floor area it divides by --- can
then be targeted rather than blanket. Where standard-setters assert
and survey the gap \citep{crrem2023,pcaf2023guidance,unepfi2022} and
industry counts its top-of-funnel rate
\citep{ftserussell2024,kpmgepra2023,kpmgepra2025,green2025,nareit2026},
this measurement carries public disclosure gate by gate to the named
curve and prices the denominator itself.

\begin{figure}[!htbp]\centering
\includegraphics[width=.82\linewidth]{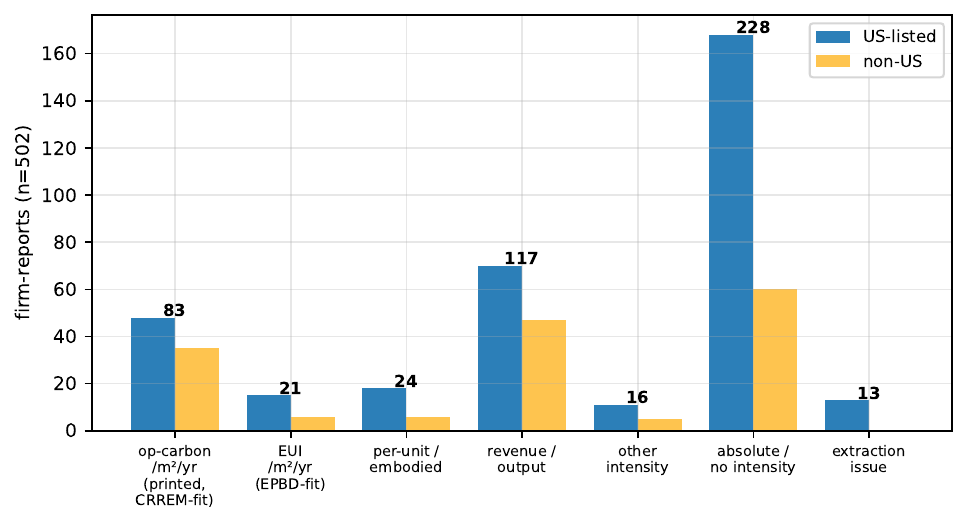}
\caption{Disclosure-class composition of the full census ($n{=}502$),
US-listed versus non-US: printed operational carbon per m$^2$/yr
(83) · energy-only EUI per m$^2$/yr (21) · per-unit/embodied (24) ·
revenue/output-normalised (117) · other intensity (16; disclosed
intensities whose denominator matches no class --- reported as a
residual, not folded into a substantive class) · absolute or none
(228, including the 7 derivable scope-1+2$\div$floor-area reports) ·
extraction issues (13). Table~\ref{tab:gap} gives the shares.}
\label{fig:gap}
\end{figure}

\begin{table}[!htbp]\centering
\caption{Disclosure-class disaggregation of the census (502 drawn
reports; 494 cleanly extracted; classifier reliability
$\kappa{=}0.95$ on an independently re-scored subset). CRREM-fit =
suitable for the CRREM operational-carbon stranding workflow;
EPBD-fit = suitable for the EU EPBD-recast EUI / energy-performance
schema. Shares are over the 494 clean rows.}
\label{tab:gap}
\small
\setlength{\tabcolsep}{4pt}
\begin{tabular}{lcccc}
\toprule
Disclosure class & n & \% & CRREM-fit & EPBD-fit \\
\midrule
operational carbon per m$^2$/yr (printed) & 83 & 16.8 & yes (op.) & yes \\
\quad memo: derivable as scope\,1+2\,$\div$\,floor area & 7 & 1.4 & derived only & --- \\
EUI per m$^2$/yr (energy only)  & 21 &  4.3 & no        & yes (EUI) \\
per-unit / per-home (embodied)  & 24 &  4.9 & no (emb.) & no \\
revenue/output-normalised       & 117 & 23.7 & no       & no \\
other intensity (unclassified denominator) & 16 & 3.2 & no & no \\
Scope 1+2 only, no intensity    & 91 & 18.4 & no        & no \\
no usable intensity             & 130 & 26.3 & no       & no \\
implausible extraction          & 5 & 1.0 & n/a         & n/a \\
\bottomrule
\end{tabular}
\normalsize
\end{table}

\subsection{A built-environment stranding-readiness sub-study (real-estate scale-up)}\label{sec:readiness}
We score a 519-report real-estate scale-up corpus and re-run the
disclosure-fitness classifier. This corpus comprises every real-estate-sector report in a
purpose-assembled supplementary collection of 1{,}235 sustainability
reports from 266 firms (fully inventoried in the data release
described under Data availability): 519 reports from 101 real-estate
firms (after collapsing report-level identifiers) spanning 2007--2024,
with US, UK, continental-European, Canadian and Australian listings
and named-file entries for unlisted issuers. Five of the 266
collection firms overlap the 2{,}246-firm panel; at the report level
the two frames share 31 report-years from 2 firms (30 of them
byte-identical PDFs), and dropping them changes neither headline
materially (Section~\ref{sec:gap}). The per-m$^2$ reporting gap is far
smaller in this core sector than in the built-environment census at
large.
\textbf{43.7\% of real-estate firm-reports (227/519; firm-clustered
BCa 95\% [35\%, 53\%]) print operational carbon per m$^2$/yr}, versus
16.5\% across the built-environment census; 65--71\% for European/UK
listings against 35\% for US listings (Section~\ref{sec:gap}),
consistent with the EPRA sBPR / EPBD regime. Extraction reliability is
established on this corpus directly: an independent second extractor
(\texttt{gemini-3.5-flash}) re-extracting a stride-sampled $n{=}80$
subset reproduces the printed-per-m$^2$ classification under the
frozen classifier at \textbf{Cohen's $\kappa{=}0.97$} ($98.8\%$
agreement; printed rates $46.3\%$ vs $47.5\%$ on the subset), computed
on uncorrected extractor output. This matches the $\kappa{=}0.95$
established on the census (second extractor
\texttt{gemini-3-pro-preview}, $n{=}50$) and shows that the
categorical assignment is \emph{reproducible} across extractors. We do
not read it as validity. Both second extractors are models of the same
provider as the primary, and errors are known to correlate across LLMs,
more strongly among models sharing a provider or architecture
\citep{kim2025correlated}; a misreading common to the Gemini family
would pass this test undetected. To probe that directly, we re-ran the
identical extraction prompt through three independent non-Google
models --- DeepSeek~V4~Flash, Qwen3.6-27B and MiniMax-M2.7 (accessed
through an OpenAI-compatible gateway) --- on the same $519$-report
real-estate corpus. Each reproduces the primary's printed-per-m$^2$
classification at Cohen's $\kappa$ of $0.86$--$0.89$ ($93$--$95\%$
agreement; \ref{app:xvendor}), and the two strongest agree with each
other at $\kappa{=}0.87$. This is lower than the same-provider
$\kappa$, as expected once the shared-provider inflation is removed,
but substantial and, decisively for the concern above, robust to a
Gemini-family artefact; on reports where two independent vendors both
return a per-m$^2$ intensity, $82$--$89\%$ of the paired values agree
within $5\%$, so the number itself, not only its presence, survives a
change of vendor. Validity of the extraction rests, beyond this
cross-vendor reproducibility, on two further checks that do not share
the Gemini failure mode: literal
grounding against the source PDF, which floors hallucination, and the
blinded human recall audit below, which tests absence. One of the two
reproducibility arms is moreover no longer re-runnable:
\texttt{gemini-3-pro-preview} was retired by the vendor mid-project
(HTTP 404, verified 9 August 2026), whereas literal grounding can be
re-run by anyone holding the PDFs. It is a categorical extraction (which denominator
is reported), so it is insensitive to the Flash tier's known upward
bias on the \emph{magnitude} of the B-score
(Section~\ref{sec:reliability}). The extracted \emph{numeric} values
also agree across the two extractors: on the subset reports where both
return a GHG intensity, $91\%$ of paired values match exactly and
$96\%$ agree within $5\%$ (\ref{app:crrem}). Cross-extractor
agreement bounds transcription error between models; it cannot test
recall on reports classified as non-disclosing, which is why a
blinded hand audit of the non-disclosing stratum (\AudN{} stratified
reports, codebook and worksheets in the released repository) is the
pre-registered gate this measurement must pass before the headline is
treated as settled
(Section~\ref{sec:limitations}).

Of the printed-or-derived per-m$^2$ reports, those dated before 2015
fall outside the pathway window, leaving \textbf{218
CRREM-constructible firm-reports} (10 of them derived rather than
printed; 76 dated 2015--2019 are scored against the clamped 2020
pathway start and flagged as such). We join them to the CRREM Global
Pathways V2.04 1.5\,$^\circ$C curves \citep{crrem2025pathways} (the
August 2025 release; the carbon curves are unchanged in CRREM's
open-access v2.05 Library, see \ref{app:crrem}): exact
country$\times$property-type curves for 58 reports, country means for
17, US/Canada property-type means (averaged over climate zones) for
77, the US/Canada all-property mean for 13, and the global curve for
53. CRREM itself retired the term ``stranding year'' in July 2025 in
favour of ``CRREM Misalignment Year'' --- the first year in which a
building's or portfolio's carbon or energy intensity is projected to
exceed the CRREM pathway --- because the older term implied a
financial outcome the pathways do not predict
\citep{crrem2025misalignment,crrem2026misalignment}; we adopt CRREM's
term for the pathway-crossing year and keep ``stranding'' only for the
generic transition-risk concept that the regulation invokes. As a
\emph{disclosure-derived stranding-readiness screen}, and
\emph{not} an asset-level stranding assessment: on the \textbf{primary,
unclamped FY$\geq$2020 subset, 35.9\% (51/142) of constructible
reports sit above the 1.5\,$^\circ$C pathway} at their reporting year;
pooling in the clamped 2015--2019 block (52.6\% above its 2020
starting value) gives 41.7\% (91/218). We quote the counts, not a
mean overshoot --- the mean is dominated by single rows and its sign
is not robust to defensible parser choices (median overshoot
$-5.4$\,kgCO$_2$e\,m$^{-2}$yr$^{-1}$). A stranding-by-2030 projection
under a flat-performance assumption is confined, with its caveats, to
\ref{app:crrem}. For US and Canadian portfolios the screen must be
read with the identification band of Section~\ref{sec:grid} attached:
placing the 90 US/CAN-tier reports at their most lenient and most
stringent climate-zone limits moves the pooled share from 26.1\% to
62.8\% around the 41.7\% zone-mean figure.

Credibility does not predict the screen. On pure-play
single-property-type REITs the association between the credibility
spine and overshoot is a null both ways: raw Spearman
$\rho{=}{-}0.15$ ($p{=}0.087$, $n{=}134$) and size-controlled partial
$r{=}{-}0.16$ on the floor-area-disclosing subset ($n{=}51$; log GFA
control) --- neither significant, and the estimate is not materially
changed by the size control. We therefore report
stranding-readiness as a disclosure-derived screen and \emph{make no
claim} that credibility predicts physical stranding. Even the
descriptive screen carries two caveats: a firm-average intensity
against one national curve is a first-order approximation of a
heterogeneous portfolio, and only the 58 exact-curve reports carry a
property-type$\times$country match.

\subsection{The grid non-identification result}\label{sec:grid}
CRREM's North American pathways are constructed on a grid of
electricity subregions crossed with climate zones: ``climate zones are
a key driver of energy consumption, whereas eGRID affects the
emissions resulting from energy consumption'' \citep[p.~22]{lbnl2024crrem}.
The published US grid has 57 such subregions, which ``correspond to
each unique combination of eGRID region and CBECS climate zone and
have a resulting CRREM pathway''; because users ``may not know their
asset's climate zone and eGRID region'', LBNL ``mapped each U.S. ZIP
code to its respective combination'' so that a pathway is retrieved
``by simply entering their asset's ZIP code'' \citep[p.~22]{lbnl2024crrem}. Holding the
(property type $\times$ climate zone) cell fixed and varying only the
electricity subregion, the V2.04 \emph{energy} limit is identical in
210 of 210 cells with two or more subregions, while the \emph{carbon}
limit varies by a median of $1.40\times$ and up to $2.79\times$. Two
identical buildings in the same climate are held to carbon targets
that differ by nearly a factor of three according to the grid they sit
on --- and no report in either corpus discloses its electricity
subregion. The consequence is scoped to V2.04 but structural: \emph{a
US portfolio's carbon verdict is not identified from disclosure, even
where the intensity is printed; its energy verdict would be.} This is
the mechanism behind the zone band above, and it makes the
energy-side comparison the robust twin of the carbon one. The
invariance is structural, not incidental: CRREM's own methodology
states that its EUI pathways ``are derived from, and designed to
complement, the CRREM Carbon Pathways'', the energy limit being
obtained ``by dividing the annual carbon-intensity values of the Carbon
Pathway by the corresponding Building-Stock Weighted Emissions Factors
for the relevant property type and jurisdiction''
\citep{crrem2026euiexplainer}. The two limits therefore differ by
exactly the jurisdiction's grid factor, which enters the carbon
verdict and is divided out of the energy verdict, so an undisclosed
subregion moves one and not the other. The energy
channel is comparably wide --- 18.3\% of census and 43.5\% of
real-estate reports print a qualifying per-area energy intensity
(EUI), largely the same reports that print carbon --- and it attrites
through the same jurisdiction (57 / 163 reports) and property-type
(37 / 137) gates, but where it exists its verdict does not move with
an undisclosed subregion, and the EPBD's national trajectories are
defined on exactly this unit \citep{eu2024epbd}.

\section{Discussion}
\subsection{Why the gap exists}
The reporting gap is not noise; it is most consistent with a
structural equilibrium. Corporate
sustainability reports are entity-level, investor-facing documents:
their reporting boundary, materiality logic, and investor-relations
framing all push towards a financially legible denominator (revenue or
units of output), not gross floor area. Floor area is an asset-operational
quantity that entity-level reporting does not naturally surface. No
widely binding mandate currently forces a per-m$^2$ denominator into
the regulated disclosure schema. On this reading, the privately optimal choice (a
revenue-normalised intensity) diverges from the socially
required one (a floor-area intensity that maps to a science-based
pathway). A
credibility instrument alone, however reliable, is therefore
insufficient for stranding triage: even a perfectly credible disclosure
is unusable for stranding if it is normalised on the wrong base.
Measuring the fitness gap addresses the half of the problem that the
credibility literature has left unexamined.

\subsection{Operational versus embodied carbon}
The fitness criterion is deliberately \emph{operational}: CRREM and the
EPBD-recast schema govern in-use carbon intensity per m$^2$/yr, so a firm
reporting only embodied or whole-life carbon, or a per-unit/per-home
denominator, is correctly counted as not operational-CRREM-fit. The criterion defines scope; it does not judge quality. The
distinction matters for how the gap should be read. For homebuilders and developers, decarbonisation leverage lies
disproportionately in the \emph{embodied} carbon of new construction
(cement, steel, structural fabric), a stream projected to remain far
off a 1.5\,$^\circ$C trajectory even under aggressive material
efficiency \citep{zhong2021materials}. For these firms a per-dwelling
embodied-intensity denominator is the architecturally correct unit
(RIBA~2030 Climate Challenge; EN~15978 whole-life assessment; RICS
embodied-carbon guidance). On an architecturally complete reading, the $\sim$4\%
of reports using it are embodied-fit rather than reporting failures. The headline
gap therefore measures fitness for the operational-stranding workflow
specifically; a complete built-environment audit would pair it with an
embodied-carbon fitness axis, which the present per-m$^2$ operational lens
does not capture and which is the natural extension for the
new-construction segment.

\subsection{Positioning}
BeDA differs from prior work in two ways. Methodologically, it inverts
the usual order: reliability is established \emph{before} the score is
used, and validity is treated as a multi-criterion question (LSEG-ESG
and RepRisk under confound controls; SBTi as a raw sectoral
corroboration) rather than a single raw correlation. The head-to-head against human coders is reported as a measurement
comparison rather than a ``treat experts as truth'' validation. Substantively, where the built-environment standards
literature \citep{crrem2023,pcaf2023guidance} and certification-adoption
work \citep{xu2025leed} establish what good disclosure should look like,
BeDA quantifies at panel scale how far the self-disclosed majority falls
from that bar.

\subsection{Implications for cities and owners}
Our results bear directly on whether building-stock decarbonisation
mandates can be enforced beyond the individual building. The obstacle
is a \emph{data} gap at source: no regulator or owner can strand-test a
building portfolio whose intensity is not disclosed on the floor-area
basis the pathway requires. For an urban regulator, the usable output is therefore portfolio triage: use the reliability-validated spine to rank disclosures for scarce audit attention, and the fitness classifier to scope a narrowly targeted disclosure mandate (require the per-m$^2$ denominator where it is missing). A portfolio owner reads the same two signals as a
transition-risk screen. The jurisdictional gradient (Europe roughly
double the US) is consistent with the gap being a closable policy choice,
not a structural feature of how buildings emit. The gap can also be
monitored automatically over time, making it possible to track whether a
new mandate is closing it.

The second thing a city needs is a link from owners to the jurisdictions that regulate them, and that is the measurement we could not yet make by machine. A registered extraction matched the markets each real-estate owner names to the sixteen enacted US building performance standards. It failed its blinded hand check on all four registered criteria (Section~\ref{sec:limitations}). The failure ran one way. The extraction counted no market that was not there, but \GtwoNoneNamed{} of the \GtwoNoneNamedN{} reports it scored as naming no market do name one, in property schedules, maps and certification tables. Every disagreement about exposure was a jurisdiction the readers found and the machine missed. For a city this means that reading corporate documents by machine undercounts the listed owners it contains. The owner has to write the jurisdiction into the disclosure; the reader cannot infer it. So we made the join by hand (\texttt{SPEC\_CITY\_BRIDGE\_V2}): each owner's latest report, coded by one author under the same codebook, screened by the same blind model coders and blind-adjudicated by another author. Of \BrOwners{} owners, \BrNamed{} name a city, metro or state market and \BrExposed{} name one inside an enacted standard, across \BrJuris{} jurisdictions; of the \BrStatusKnown{} with a 2020 or later report and a usable intensity read, \BrMetricMatch{} print the jurisdiction's metric with a property type and \BrReadable{} also disclose the floor area. These counts are \BrStatus. A design-weighted reading of the forty hand-checked reports (\texttt{SPEC\_EXPOSURE\_ESTIMATE}) gives the same picture at the level of reports: \ExpoShare{} (95\% interval \ExpoLo{} to \ExpoHi{}) name a market inside an enacted jurisdiction. A city can name most of its listed owners from what they publish, and can read almost none of them.

\subsection{Limitations}\label{sec:limitations}
The validity claim is anchored on \emph{convergent} external criteria
(LSEG-ESG, SBTi, RepRisk), not on a conventional gold-set validation.
The head-to-head against a two-coder human pass at $n=100$ found the
human coding's discriminative range too compressed (stdev $2.73$ vs the
LLM's $9.99$ on the $0$--$32$ scale) to anchor construct validity in
the conventional sense, despite high inter-coder agreement
($w\kappa{=}0.857$ on B). We interpret the divergence as human leniency
but cannot exclude the competing hypothesis that the spine
over-discriminates on compliance-correlated surface features (document
length, table density, presence of a GRI index). The three-criterion
convergence favours our interpretation, since a purely surface-feature
score would be unlikely to track SBTi validation, RepRisk incidents, and
LSEG-ESG simultaneously. However, resourcing-correlated surface features
may co-vary with all three. The stratified-extreme adjudication re-code
is the test we have prepared to settle the question (selection and
blinded protocol included in the code release described under Data
availability). The instrument is
likewise tied to a single operationalisation: all validators in the
reliability battery score against the same fixed SSSR rubric text and
identical prompt (\ref{app:implementation}), so cross-model concordance
does not test robustness to rubric re-wording.

The confound set has known boundaries: disclosure volume is
operationalised as log report length, a text-centric proxy that does
not capture visual or tabular verbosity in these highly multimodal
reports. There is also no direct balance-sheet size control (no log
assets or market capitalisation): firm size is proxied only indirectly
through the sector, listing-region and volume terms, and incident
counts may additionally scale with media coverage and firm
prominence. Adding a financial
size control is a natural extension once coverage permits. Report
language is not modelled either: the panel inherits the parent corpus's
collection of published report versions, and any covariation between
listing region and the language or version collected therefore enters
the regional comparison uncontrolled.

The leakage
probe bounds \emph{both} input channels (text $r=0.886$, vision
$r=0.826$, $n=30$) and leaves a small systematic identity component on the
text channel and a modest page-sampling sensitivity on the vision
channel, neither consistent with strong firm-identity recognition. At
$n=30$ the probe bounds average behaviour; it is underpowered to rule
out memorisation for individual high-profile firms.

The CRREM analysis is a disclosure-derived stranding-\emph{readiness}
screen, not an asset-level stranding assessment: a firm-average
intensity against one national pathway is a first-order approximation of
a heterogeneous portfolio, exact property-type$\times$country curves
are available only for the European reports, and for US/Canadian
portfolios the carbon verdict is bounded, not identified
(Section~\ref{sec:grid}). We deliberately make no predictive claim:
the credibility--overshoot association is a null with and without the
size control (Section~\ref{sec:readiness}). The numeric intensities
feeding the screen are checked by cross-extractor agreement and
literal grounding. The first is a reproducibility bound, not a
validity test: the same-provider extractor pairs are Gemini models,
and correlated errors across LLMs are documented
\citep{kim2025correlated}. We therefore added a cross-vendor extraction
arm --- three independent non-Google models reproducing the per-m$^2$
classification at $\kappa{=}0.86$--$0.89$ (Section~\ref{sec:reliability},
\ref{app:xvendor}) --- which bounds a Gemini-family artefact, though
agreement across all four vendors still cannot certify an error common
to every model. The second floors hallucination. Neither can test
\emph{recall}: a report classified as non-disclosing that
in fact discloses would inflate the headline, and the deep audits that
produced this revision found extraction errors of exactly that
family (all corrected and documented in the released corrections
file). The blinded hand audit of \AudN{} stratified non-disclosing reports,
packaged with a written codebook and a conservative-reading rule
(ambiguity codes \emph{against} our thesis), is therefore a
pre-registered gate: if recall on that stratum is below 90\%, the
headline moves. The coding of record is one author's, and we report no inter-coder
reliability. Around it we placed three checks, specified and
hash-signed before any model was called. A machine disagreement
screen: three text-only language models from independent vendors and
one PDF-reading model coded every report blind under the codebook
verbatim, and a row kept the human code without a second human read
only when the text-model majority and the PDF reader both agreed with
it (\ScrReleasedA{} rows); every other row, every report with a
deficient text layer and a seeded random \ScrAuditA{} of the released
rows went to a third author, who adjudicated from a packet showing no
code and no reason for a row's inclusion. A blind test--retest by the
first coder on a seeded subsample. And rulings, made before the run
and against our thesis, on five configurations the codebook does not
settle. On the final codes, \GoneSentence{}. The PDF reader reproduced
\ScrPdfHitYA{} of the human coder's positive codes and \ScrPdfHitNA{}
of the negative codes; that is reproducibility of the coding, not
reliability, and no model's code became a final code. The
\AudFieldN{}-report field pass found the extracted value correct in
\AudFieldValue{} reports and the floor area in \AudFieldGfa{};
\AudFieldGfaMissed{} of the floor-area errors are a stated floor area
the extractor left blank, so the floor-area gate is a lower bound. Both frames' PDF stores are held
in full: all 519 Frame-B documents are inventoried in the collection
manifest and present as files alongside the 502 census reports, so
both extractions are re-runnable end-to-end (the PDFs are distributed
as document identifiers rather than files only for licensing reasons;
see Data availability). A direct Frame-B recall audit is therefore
feasible rather than inherited from the census measurement: a blinded
Frame-B hand-check package --- a \ScrMktN{}-report market-naming
worksheet for the city bridge and the Frame-B non-disclosing stratum
for recall --- was built alongside the census package under the same
codebook rules and coded, screened and adjudicated the same way. On
the first coder's block the market check fails all four registered
criteria (name precision \GtwoPrecision{} against 90\%; \GtwoNoneNamed{}
of the reports scored as naming no market do name one; \GtwoHq{}
headquarters-as-market errors; exposure agreement \GtwoExposure{}
against 90\%), every exposure disagreement running the same way: the
coder finds an enacted-standard jurisdiction the extractor missed.
\GtwoSentence. Under the registered rule a failed check means no
city-bridge number is quoted as a finding.

The built-environment lead stratum for the
confound-controlled \emph{validity} tests remains thin in $n$ (the new
real-estate firms lack the subscription LSEG/RepRisk coverage), so the
full panel carries that inferential weight. The ClimateBERT comparator is
run zero-shot rather than fine-tuned, for the label-availability reason
given in Section~\ref{sec:baselines}; its scores therefore bound what a
published off-the-shelf classifier achieves, not what a task-tuned one
might. All rates are, moreover, conditional on the sampling frame: the
panel covers publicly listed firms that publish sustainability
reports, so the fitness rates characterise reporting listed firms and
should not be generalised to unlisted or non-reporting building
owners. Scoring is single-pass at a low but non-zero temperature, and
P4 extraction uses provider-default decoding
(\ref{app:implementation}); within-model run-to-run variation is not
separately quantified.
None of these undermine the disclosure-fitness gap measurement or the
instrument's reliability, the latter established at $n=407$ and
replicating across the Gemini and Gemma families and across modality
(text-only Pro arm), with the caveat noted in
Section~\ref{sec:reliability} that the panel-scale scoring validators are all
Google models: the pre-registered cross-vendor arm bounds a vendor
artefact on the scoring instrument only narrowly ($r{=}0.725$ against
a $0.70$ threshold, on an extreme-groups subset). The extraction
pipeline now carries its own cross-vendor arm --- three non-Google
models reproducing the per-m$^2$ classification at $\kappa{=}0.86$--$0.89$
(\ref{app:xvendor}) --- so the extraction $\kappa$ is no longer
Gemini-versus-Gemini, though no agreement statistic can certify an error
shared by every vendor.

\section{Conclusions and policy implications}
Using BeDA, a reliability-validated multimodal LLM auditor, we measure
at panel scale whether building-decarbonisation disclosure can be
carried to the science-based pathways that stranding regulation
assumes. Mostly it cannot. One census firm-report in six prints the floor-area-normalised intensity, one in fifteen reaches a property-type pathway, one in twenty discloses its own denominator, and most failures come from reports that lump the portfolio together, not from gaps in the pathway library. The European rate is
roughly twice the US rate, and the conditioning evidence locates the
lever in the benchmark a firm reports into rather than its
jurisdiction. For US portfolios the carbon verdict is, in any case, not identified from disclosure, because the pathway's carbon limit moves with an electricity subregion no report names.

The gap is not neutral across owners. A firm's portfolio scope barely changes across reports, so the tests are firm-level (majority scope; a firm passes if any qualifying report passes). Every single-country owner with a printing, in-window report passes the jurisdiction gate ($42$ of $42$), against $6$ of $17$ multi-country owners (Fisher exact $P=4\times10^{-8}$; property-type gate $88.1\%$
versus $29.4\%$, $P=2\times10^{-5}$). Part of this is the gate's
construction --- a portfolio spanning jurisdictions cannot resolve to
one --- but the two variables come from independent extraction passes,
and the incidence is the point: the owners whose buildings span the
most covered cities are precisely the ones no single pathway can read.
Printing, denominator disclosure and both report-length comparisons
cannot be distinguished at firm level at this sample size and are not
read as findings; if anything, multi-country owners disclose the
floor-area denominator \emph{more} often, the opposite of a
resourcing story.

The prescription follows from the measurement, and the unit of account
it implies is the one building performance standards are converging on:
Local Law~97's
implementing rule requires per-building intensity by occupancy group
with the floor area of each \citep{nycdob2026rule}; Boston's
portfolio compliance route --- the portfolio-basis mechanism the
tradable-BPS literature records as the exception among US standards
\citep{spiegelfeld2022tradable} --- retains per-building reporting
\citep{boston2021berdo}; Maryland's standard fixes net direct emissions
in kg\,CO$_2$e per square foot by property type for every commercial or
multifamily building of 35,000~ft$^2$ or more \citep{md2022csna,mdcomar2628};
and Colorado's Regulation~28 sets greenhouse-gas-intensity targets in
kg\,CO$_2$e per square foot (or site EUI in kBtu per square foot) by
property type for buildings of 50,000~ft$^2$ or more \citep{co2021hb1286,coreg28}. What company-level disclosure rules lack is that same unit: intensity split by asset type and jurisdiction, reported with the floor area it divides by. That requirement is
cheaper than a broad disclosure mandate, and this instrument can
monitor whether it is being met.

Several extensions remain open: the stratified-extreme adjudication re-coding
prepared to probe the human-leniency pattern documented in
\ref{app:goldset}, an embodied-carbon fitness axis for the
new-construction segment, and panel-scale extraction quality control
with supervised deep-learning comparators. One extension we investigated and decline on the data: a regulatory event study of the per-m$^2$ rate around the EPBD recast or NFRD/SFRD dates cannot be identified on this panel. The corpus contains no continental-European \emph{listings} among the census exchanges, the treated group shrinks to nine UK firms with enough reporting years, and the EU--US gap is a level difference present in the first observed year --- a pre-trend the design cannot survive.

\section*{Changes from the arXiv v1 preprint}
This version supersedes arXiv:2607.22006v1 (24 July 2026) after an
internal defect audit and re-founding. The measurement moved from a
200-report stratified sample to the 502-report census under a written
specification (printed-only basis; pre-committed diversified-portfolio
rule; documented data corrections); the sample-based headline
(21.5\%, Wilson interval) is superseded by the census figure (16.5\%,
firm-clustered model-based interval), the ``design-based'' and
frame-weighted-19.6\% language is withdrawn (the recency selection had
no inclusion probabilities), the real-estate rate moves from 45.5\% to
43.7\% (printed-only, corrected chain), the stranding screen from
39\%/215 to the FY$\geq$2020 primary 35.9\% with the pooled 41.7\%/218
and the zone identification band attached, the mean overshoot is
withdrawn as single-row-sensitive, and the credibility--overshoot
attenuation story is restated as a null. Of the headline quantities
that moved, most moved \emph{against} the previously published numbers
and in favour of the paper's thesis, while every extraction-layer
correction (restored tonne spellings, area qualifiers, annualisation)
moved \emph{toward more disclosure}; the corrections were not steered.
The full old-to-new ledger, the measurement specification, and the
claims registry that regenerates every number in this paper ship with
the repository.

\section*{CRediT authorship contribution statement}
\textbf{Jingyi Xu:} Conceptualization, Methodology (governance),
Investigation (blinded audit coding), Formal analysis (governance),
Writing -- original draft, Writing -- review \& editing.
\textbf{Minghui Cheng:} Formal analysis (benchmarking), Writing --
review \& editing.
\textbf{Anchen Sun:} Conceptualization (lead), Methodology (computational),
Software, Investigation (audit adjudication), Validation, Writing --
original draft, Writing -- review \& editing.

\section*{Declaration of competing interest}
Anchen Sun is an employee of Google; the Gemini and Gemma models used
as scoring backbones in this study are developed by Google. The
authors declare no other competing financial or personal interests.

\section*{Data availability}
The analysis code (scoring, reliability, confound-controlled validity,
disclosure-fitness classification, and the CRREM construction) and the
derived, de-identified score-level data (per-report dimension scores,
disclosure-fitness classes, per-report cascade gate status, and CRREM
stranding-readiness outcomes sufficient to reproduce every figure and
table) will be released in a public repository (GitHub, archived on
Zenodo with a DOI) upon publication, together with the measurement
specification, the data-corrections file with evidential bases, the
claims registry, the frozen unit-string fixture, and the recall-audit
codebook and worksheets. The CRREM Global Pathways V2.04 dataset (August 2025) is the
analysed release; CRREM now publishes its pathway data open-access
through the CRREM Library (v2.05), and both are used here under
CRREM's terms of use with attribution; the underlying
corporate report PDFs remain subject to their original source
terms and are therefore provided as document identifiers rather than
files. The Loughran--McDonald master dictionary and the SBTi public
dashboard are available from their original providers; the ClimateBERT
baseline uses public Hugging Face checkpoints whose resolved revision
hashes are recorded in the same release; LSEG-ESG and
RepRisk are subscription data and cannot be redistributed.

\section*{Funding}
This research did not receive any specific grant from funding agencies in
the public, commercial, or not-for-profit sectors.

\section*{Acknowledgements}
We thank Ziyuan Xia (PhD candidate) and Professor Saixing Zeng at the
Antai College of Economics and Management, Shanghai Jiao Tong University
(SJTU) for developing the SSSR framework and corpus on which this study
builds, and for producing the independent two-coder human gold set. We acknowledge the Carbon Risk Real Estate
Monitor (CRREM) for the Global Pathways data used under
academic-licence terms.

\clearpage
\appendix

\section{Instrument reliability and identity-anonymisation leakage}\label{app:reliability}
The full cross-model and cross-\emph{family} concordance matrix underlying
Section~\ref{sec:reliability} is given in Table~\ref{tab:xfam} (Pearson $r$
against the primary \texttt{gemini-3.1-pro-preview}; the
standards-compliance spine (B) is the credentialed dimension; column
labels are short identifiers, full model identifiers in
Table~\ref{tab:models}).

\begin{table}[!htbp]\centering
\caption{Cross-model and cross-\emph{family} concordance (Pearson $r$;
reference \texttt{gemini-3.1-pro-preview}). The within-Pro multimodal
column is computed on the full common-scored set ($n=495$); all other
columns use the shared audit subset ($n=100$, except $n=99$ for
\texttt{gemini-3.5-flash}).}
\label{tab:xfam}
\small
\setlength{\tabcolsep}{4pt}
\begin{tabular}{lcccccc}
\toprule
SSSR dim. & 3-pro & 3-pro & 3-flash & 3.5-flash & gemma-4 & gemma-4\\
     & (mm)  & (txt) & (mm)    & (mm)      & 31B (txt) & 26B (txt) \\
\midrule
B (standards compliance)  & 0.844 & 0.847 & 0.760 & 0.744 & 0.819 & 0.764\\
A (topic breadth)         & 0.562 & 0.460 & 0.572 & 0.473 & 0.376 & 0.453\\
C (multimodal evidence)   & 0.384 & 0.316 & 0.531 & 0.449 & 0.186 & 0.194\\
\bottomrule
\end{tabular}
\normalsize
\end{table}

\subsection*{Identity-anonymisation leakage probe}\label{app:leak}
\begin{figure}[!htbp]\centering
\includegraphics[width=.6\linewidth]{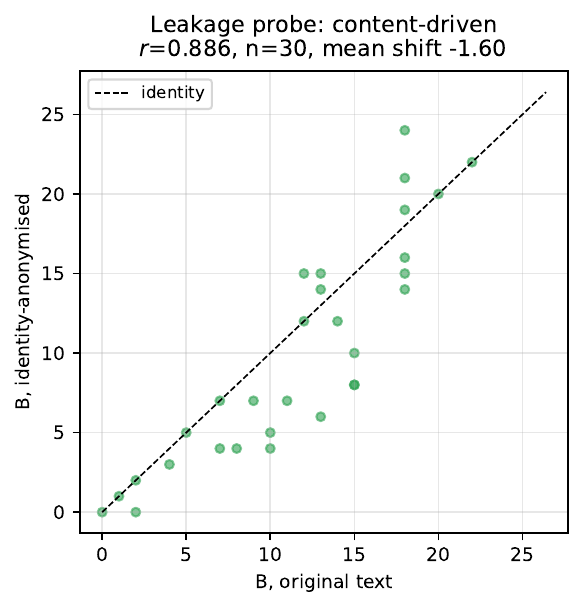}
\caption{Text-channel identity-anonymised re-scoring ($r=0.886$, shift
$-1.6$). The vision-channel probe (cover-page-omitted multimodal,
$r=0.826$, shift $+2.17$) is summarised in the text and supports the
spine being leakage-robust on both channels.}
\label{fig:leak}
\end{figure}

We probe both channels of the input independently. Re-scoring 30 reports
with firm identifiers removed from the \emph{text channel} by automated
pattern replacement (ticker and firm-name tokens replaced with a
neutral placeholder; mechanics in \ref{app:implementation}),
model held
constant, yields $r=0.886$ between original and anonymised B-scores,
with a small systematic shift ($-1.6$ on the 0--32 scale)
(Fig.~\ref{fig:leak}). The text-channel component of the spine is
therefore predominantly content-driven and not firm-reputation-driven,
with a small residual identity component. To bound the \emph{visual}
channel we re-score the same 30 reports with the full multimodal
pipeline in both arms; the only difference is the cover page (page~0,
where corporate logos, branded chart styles and recognisable building
photographs typically live). In the anonymised arm the cover page is
omitted from the visual rendering and three body pages are sampled
instead. This yields $r=0.826$ between original and cover-omitted
multimodal B-scores, with a small \emph{positive} shift of $+2.17$,
the opposite direction from what cover-page identity recognition
would predict. The likely mechanism is selection: omitting the cover
page concentrates the visual context on body pages where substantive
compliance content (GRI content-index tables, assurance statements,
methodology notes) is disclosed more densely than on the cover, modestly
raising B on average. Taken together, the two probes \emph{bound both
channels}: the spine is leakage-robust and is not driven by cover-page
identity recognition; visual-channel page sampling does measurably affect
scores (${\approx}{+}2$ B on a $0$--$32$ scale), reported as a separate
sensitivity rather than folded into a single ``leakage'' claim.

\section{An LLM-versus-human measurement-divergence study}\label{app:goldset}
A two-coder human gold set was independently produced by a domain
collaborator (SJTU; 100 stratified non-overlapping reports across six
sectors, $50$ pre-/$50$ post-2018; both coders applied the scoring
anchors of the rubric package). Inter-coder reliability is high in
every dimension and almost perfect for the credibility spine:
$\mathrm{B}\,w\kappa{=}0.857$, $\mathrm{A}\,w\kappa{=}0.687$,
$\mathrm{C_1}\,w\kappa{=}0.658$, $\mathrm{C_2}\,w\kappa{=}0.750$,
$\mathrm{C_4}\,w\kappa{=}0.836$. By the Landis--Koch benchmarks
\citep{landiskoch1977}, the gold set is internally reliable. We then compare the LLM B-score against the mean human
B-score under two input regimes: the restricted-input cross-model arm
(4{,}000-character text excerpt + three page renders) and a separately
re-scored full-PDF arm ($n=96$ of $100$ successfully re-scored). Both
arms
yield $r{\approx}{-}0.24$--${-}0.29$ LLM-vs-mean-human on B, while the
two LLM arms themselves agree closely ($r{=}0.811$,
$w\kappa{=}0.794$): the negative LLM--human correlation is therefore
\emph{not} an input-restriction artefact.

\begin{figure}[!htbp]\centering
\includegraphics[width=.62\linewidth]{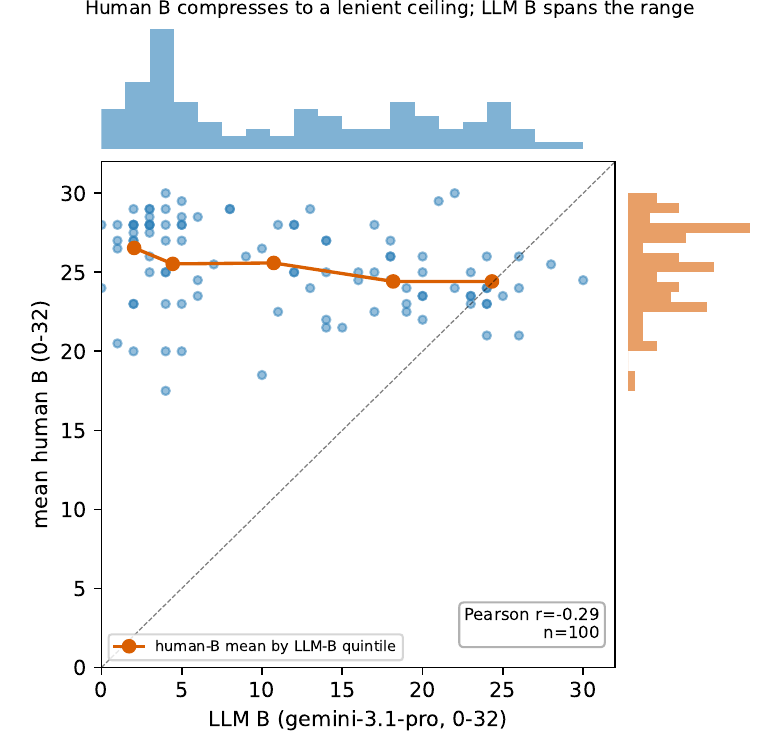}
\caption{The human gold set compresses to a lenient ceiling while the
spine spans the scale. Each point is one of the 100 gold-set reports
(LLM B on $x$, mean two-coder human B on $y$), with marginal histograms;
the orange line is the mean human B within each LLM B-quintile. Humans
vary by $2.1$ points across the LLM's $22.3$-point range (Pearson
$r{=}{-}0.29$). This compression, together with a weak downward drift in human
scores across LLM quintiles, drives the negative LLM--human
correlation: a drift, not a substantive reversal.}
\label{fig:leniency}
\end{figure}

The divergence is mechanical in form, and the leniency-compression
pattern that accompanies it is directly observable.
Binning the 100 reports by LLM B-quintile (Fig.~\ref{fig:leniency}), the
mean human B is essentially flat: $26.5,\,25.5,\,25.6,\,24.4,\,24.4$
from the lowest to the highest LLM quintile, a spread of $2.1$ points
against the LLM's $22.3$-point spread. Reports that the LLM scores near
$2/32$ still receive ${\approx}26/32$ from the human coders. The two coders are
reliable (B $w\kappa{=}0.857$) but exercise only the upper ${\approx}$third
of the scale (stdev $2.73$, range $18$--$30$), whereas the LLM spans it
(stdev $9.99$, range $2$--$32$). The pattern recurs \emph{across
dimensions}: LLM--human agreement is positive only on the one dimension
where the human pass uses a wide range (visual evidence $\mathrm{C_1}$,
$r{=}{+}0.29$) and turns negative on the compressed upper-band dimensions
(A $r{=}{-}0.18$; B $r{=}{-}0.29$). We interpret this as a
leniency / central-tendency regime. We advance this as a hypothesis;
the mechanism is not established, and Section~\ref{sec:limitations} states the competing
interpretation (LLM over-sensitivity to compliance-correlated surface
features) we cannot exclude.

The divergence is bounded by two further results. First, it is not an artefact of \emph{our} model or prompt. An
independent LLM pipeline (the parent SSSR cross-sector auditor, which
uses a different scoring model on a $0$--$40$ B scale and a larger
$11{,}821$-report corpus with its own human panel
\citep{xia2025global,nan2025governance}) reproduces the sign
and approximate magnitude of the B divergence ($r{\approx}{-}0.38$,
$n{=}100$). Two pipelines differing in model, scale, prompt and corpus
both anti-correlate with expert holistic coding while each is internally
reliable. This is most consistent with the divergence arising in human
coding of long structured ESG documents. Second, the quantity the external criteria recover is the LLM
ordering, not the human ceiling: the convergent-validity tests
(Section~\ref{sec:validity}, \ref{app:validity}) are positive for the
spine. On the gold-set firms matchable to RepRisk, the spine tracks
incident counts more closely than the human pass (Spearman $0.21$ vs
$0.08$, $n{=}68$; underpowered but directionally consistent). We
therefore anchor construct validity on the external criteria and the
cross-pipeline replication, not on the human pass.

\paragraph{A third independent method: the lexicon floor} A transparent
Loughran--McDonald lexicon baseline (text-only, no learning) provides a
further check. It does not recover the $B$ construct (strongest
panel signal, negative-word fraction vs $B$, $r{=}0.21$; the cheap-talk
composite (the sum of the lexicon's uncertainty and weak-modal word
fractions) $r{\approx}0$), confirming the spine captures variance a sentiment
lexicon does not. Yet on the gold set the same lexicon \emph{aligns with
the LLM ordering, not the human ceiling}
(negative-word fraction vs LLM $B$ $r{=}{+}0.44$; vs mean-human $B$
$r{=}{-}0.10$). That a multimodal LLM, an independent LLM pipeline, and a
non-learned text lexicon all diverge from the lenient human scores in the same direction is most consistent with the
divergence arising in human coding of long structured disclosures. (The
zero-shot ClimateBERT baseline of \ref{app:validity} tracks
neither pass on the gold set ($r{=}{-}0.03$ vs the human mean,
$r{=}{-}0.30$ vs the LLM, $n{=}85$) and is therefore uninformative for
this adjudication.)

\section{Confound-controlled external validity and baseline comparison}\label{app:validity}
The confound-controlled convergent- and criterion-validity estimates
discussed in Section~\ref{sec:validity} are reported in Table~\ref{tab:valid}.
Table~\ref{tab:baseline_validity} then runs every estimable baseline of
Table~\ref{tab:baselines} through the identical battery, with all scores
z-standardised per analysis sample so the LSEG coefficient and the RepRisk
incidence-rate ratio are per~1~SD and directly comparable across rows.

\begin{table}[!htbp]\centering
\caption{Confound-controlled validity. Controls: log report length,
listing region (US/non-US), sector division, reporting year;
firm-clustered robust SEs; the partial-$r$ CI is a firm-clustered
bootstrap (2{,}000 cluster resamples; percentile interval). The
firm-level incident model uses firm-mean disclosure volume, listing
region and sector division (no reporting-year term; one observation per
firm, hence unclustered maximum-likelihood CIs). The fully controlled
$\beta$ is unstandardised (per-SD
estimates in Table~\ref{tab:baseline_validity}).}
\label{tab:valid}
\begin{tabular}{lcc}
\toprule
Specification & Estimate & 95\% CI / $p$\\
\midrule
B$\leftrightarrow$ESG, raw Pearson $r$ & 0.236 & ---\\
\quad volume-adjusted partial $r$ & 0.203 & [0.09, 0.31]\\
\quad fully controlled $\beta$ & 0.005 & [0.002, 0.007]\\
Built-env $\times$ panel interaction & 0.013 & $p{\approx}2\times10^{-6}$\\
B $\rightarrow$ incidents, NegBin IRR & 1.052 & [1.034, 1.071]\\
\bottomrule
\end{tabular}
\end{table}

\begin{table}[!htbp]\centering
\caption{Baseline validity matrix: every comparator on the identical
validity battery (LSEG and RepRisk confound-controlled, SBTi a raw
point-biserial; all scores z-standardised; LSEG $\beta$ and
RepRisk IRR per~1~SD; partial-$r$ CIs are firm-clustered bootstrap
percentile intervals from 1{,}000 cluster resamples). The cheap-talk share is negative-coded, so validity
convergence appears as inverse estimates. ClimateBERT rows use the subsample
with defined shares ($n{=}677$ firm-years for LSEG, $1{,}033$ firms for
RepRisk, $116$ firms for SBTi); other rows $n{=}700$, $1{,}052$, $119$.
ClimateBERT shares are computed over sentence-grouped chunks of at most
300 words and are defined for reports with at least five chunks the
detector classifies as climate-related.
n.s.\ $=$ not significant at 5\%.}
\label{tab:baseline_validity}
\small
\setlength{\tabcolsep}{3.5pt}
\begin{tabular}{lcccc}
\toprule
Score & LSEG partial $r$ & LSEG $\beta$/SD & RepRisk IRR/SD & SBTi $r_{pb}$\\
 & [95\% CI] & ($p$) & ($p$) & ($p$)\\
\midrule
Standards-compliance spine (B-only) & $+0.203$ [$+0.09,+0.31$] & $+0.040$ ($6{\times}10^{-5}$) & 1.410 ($6{\times}10^{-9}$) & $+0.360$ ($6{\times}10^{-5}$)\\
Spine + multimodal ($B{+}C$, z-mean) & $+0.109$ [$-0.00,+0.21$] & $+0.021$ (0.027) & 1.371 ($3{\times}10^{-9}$) & $+0.310$ (0.0006)\\
ClimateBERT specificity & $+0.007$ [$-0.11,+0.14$] & $-0.003$ (n.s.) & 1.056 (n.s.) & $+0.180$ (0.054)\\
ClimateBERT cheap-talk & $+0.014$ [$-0.10,+0.12$] & $+0.004$ (n.s.) & 0.767 ($7{\times}10^{-6}$) & $-0.206$ (0.027)\\
LM negative-word fraction & $-0.014$ [$-0.17,+0.14$] & $-0.005$ (n.s.) & 1.042 (n.s.) & $-0.050$ (n.s.)\\
\bottomrule
\end{tabular}
\normalsize
\end{table}

\begin{figure}[!htbp]\centering
\includegraphics[width=.95\linewidth]{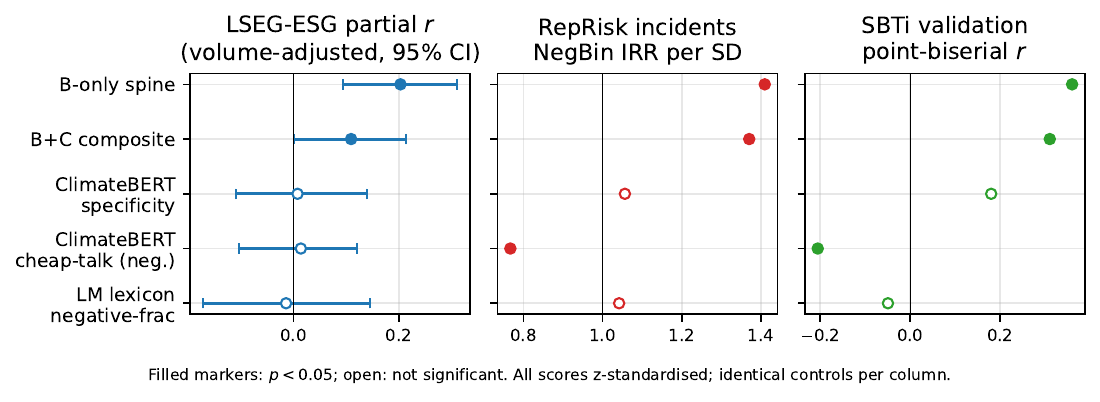}
\caption{The baseline validity matrix of Table~\ref{tab:baseline_validity}
as a dot plot. Filled markers are significant at 5\%; open markers are not.
Only the spine-based scores (standards-compliance alone and the
spine + multimodal composite) are positive and significant on all three
external criteria, the spine alone most strongly; the cheap-talk share's
two significant estimates run in the greenwashing direction
(negative-coded).}
\label{fig:baselines}
\end{figure}

The matrix supports three readings. First, the dimension ablation: adding the
multimodal evidence dimensions ($C_1$, $C_2$, $C_4$) to the spine
\emph{dilutes} every point estimate (LSEG partial $r$
falls from $+0.203$ to $+0.109$; SBTi from $+0.360$ to $+0.310$; CIs overlap),
consistent with their cross-family unreliability
(Section~\ref{sec:reliability}) and with confining substantive claims to
$B$. Second, the zero-shot ClimateBERT cascade does not recover the
spine construct (specificity share vs $B$: $r{=}{-}0.09$ panel-wide,
$n{=}6{,}714$) and carries no significant LSEG signal, but its cheap-talk
share shows a genuine signal on two of the three criteria: firms with more
commitment-without-specificity language are less likely to hold a
validated SBTi target ($r_{pb}{=}{-}0.206$) and show fewer recorded
incidents (IRR $0.767$). This is an independent, method-diverse
corroboration of the direction of association the spine shows, subject
to the same prominence caveat as Section~\ref{sec:validity}. Third, only
the spine-based scores converge across all three external criteria
simultaneously; the lexicon floor is flat throughout. On the gold set, the zero-shot
cascade tracks neither pass (specificity share $r{=}{-}0.03$ vs the human
mean, $r{=}{-}0.30$ vs the LLM, $n{=}85$), so it neither supports nor
contradicts the human-ceiling reading of \ref{app:goldset}.
Fig.~\ref{fig:baselines} shows the matrix graphically.

\section{CRREM construction and robustness}\label{app:crrem}
\paragraph{Version pin} All pathway joins use the CRREM Global
Pathways V2.04 dataset (August 2025) \citep{crrem2025pathways}, which
remains the dataset behind CRREM's own pathway explorer. CRREM has
since retired its Excel Risk Assessment Tool (1 July 2026) in favour
of an open-access Library whose v2.05 release (30 April 2026)
re-issues the Global Pathways in long format and refreshes carrier
emission factors, stating that the pathways, methodology and
underlying datasets are unchanged \citep{crrem2026retire,crrem2026library}.
We checked this directly: every 1.5\,$^\circ$C carbon series used here
--- all country$\times$property-type curves and all US/Canada
subregion$\times$climate-zone curves --- is numerically identical in
the V2.04 and v2.05 workbooks (13{,}950 and 30{,}597 matched cells,
no differences). The v2.05 workbook does not re-issue V2.04's global
all-country curve, on which the 53 global-curve reports rely, which is
why the screen is reported on V2.04 rather than re-based. CRREM's next
pathway revision (refreshed carbon and EUI pathways, a reintroduced
2\,$^\circ$C scenario, new property types) is announced for June 2027
with go-live expected in November 2027 \citep{crrem2026eui}; no result
in this paper moves with the v2.05 Library.

Each per-m$^2$ operational-carbon intensity is mapped to a
CRREM Global Pathways V2.04 1.5\,$^\circ$C curve \citep{crrem2025pathways} using the jurisdiction-appropriate
pathway: US and Canadian firms to the CRREM US/Canada national-mean curve
for their property type (sheet~6 of the published pathways workbook,
averaged over the climate-zone columns; $13$ of the $90$ US/CAN
reports use the all-property-type mean), European/UK firms to the
exact country$\times$property-type curve ($58$) or a country mean
($17$), and the remainder (mostly Australian and multi-jurisdiction)
to the global curve ($53$). Country strings are resolved by an
inspectable alias table that distinguishes a genuinely
multi-jurisdiction portfolio (for which the global curve is correct)
from a parser failure; free-text property types are normalised to
CRREM codes by longest-match against a fixed dictionary; where no
property-type curve matches, a report falls back to the all-property
mean (US/Canada) or the country mean (Europe). Intensities are
normalised to kgCO$_2$e$\cdot$m$^{-2}\cdot$yr$^{-1}$ with explicit
unit handling (t/tonne/ton$\to$kg, lb$\to$kg,
per-ft$^2\to$per-m$^2$, ``per-1000-sqft'' multipliers, and per-month
$\times12$ annualisation), a physical-plausibility band of
$[1, 2000)$, and a frozen 400-string unit-fixture regression test;
the CRREM Misalignment Year --- CRREM's term since July 2025 for what
its earlier guidance called the stranding year
\citep{crrem2025misalignment} --- is the first year $\geq$ the
reporting year at which the flat-held firm intensity exceeds the
declining pathway.

\paragraph{Temporal alignment} The V2.04 pathways span 2020--2050.
Constructibility requires a reporting year of 2015 or later. The $76$
constructible reports dated 2015--2019 are compared against the
pathway's 2020 starting value, the least stringent point of the
declining curve; each such row carries an explicit clamp flag in the
released outcome file. The unclamped FY$\geq$2020 subset is the
primary screen ($35.9\%$ above pathway, $51/142$); the clamped block
sits at $52.6\%$ above its 2020 starting value, and the pooled figure
is $41.7\%$ ($91/218$).

\paragraph{Numeric extraction agreement} The $n{=}80$ second-extractor
subset (\texttt{gemini-3.5-flash}) allows a numeric check beyond the
categorical $\kappa$, computed on uncorrected extractor output: on the
$46$ reports where both extractors return a GHG intensity, $91\%$ of
paired values match exactly, $96\%$ agree within $5\%$, and the median
absolute difference is $0.0\%$. Two reports ($4\%$) show material
disagreement, consistent with occasional misreading of complex tables
--- the error family the recall audit and field-correctness pass
(Section~\ref{sec:limitations}) are designed to bound against human
ground truth.

\paragraph{Credibility--overshoot association} On pure-play REITs the
raw Spearman $\rho$ is $-0.15$ ($p{=}0.087$, $n{=}134$) and the
size-controlled partial $r$ is $-0.16$ (floor-area-disclosing subset,
$n{=}51$, log GFA control); on the full constructible set,
$\rho{=}{-}0.11$ ($p{=}0.105$, $n{=}218$) and partial $r{=}{-}0.06$
($n{=}85$). No estimate is significant, with or without the size
control; we report stranding-readiness as a disclosure-derived screen
and make no predictive claim.

A forward projection is possible, but we deliberately keep it out of the
main text because it compounds three strong assumptions. Under a
\emph{flat-performance} assumption (firm intensity held constant against
the declining pathway), $75.7\%$ ($165/218$) of the constructible
reports would cross their CRREM pathway by 2030. This figure is a
first-order illustration only: it assumes no decarbonisation effort,
maps a firm-average intensity to a single national curve, and carries
an exact property-type$\times$country match for only the $58$
exact-curve reports. It should not be cited as a stranding forecast;
the point-in-time FY$\geq$2020 share ($35.9\%$,
Section~\ref{sec:readiness}) is the screen we stand behind, read with
the zone identification band of Section~\ref{sec:grid}.

\section{Cross-vendor extraction reliability}\label{app:xvendor}

The extraction reliability of the main text (a same-provider $\kappa$) is complemented by a
pre-registered cross-vendor arm (\texttt{SPEC\_EXTRACTION\_SWEEP.md}): the byte-identical CRREM
extraction prompt, run through three independent non-Google models on the full $519$-report
real-estate corpus. Because the gateway serves text only, we separate a modality effect (a
same-model text arm, \texttt{gemini-3.1-pro-preview} on the extracted text) from the vendor
effect. We report $\kappa$ under two classifiers: the headline parser
(the printed per-m$^2$ classification behind the $43.7\%$ headline rate) and the
coarser regex used for the same-provider $\kappa{=}0.97$.

\begin{table}[htbp]\centering\small
\caption{Cross-vendor reproducibility of the printed-per-m$^2$ classification. $\kappa$ = Cohen's
$\kappa$ against the frozen primary (\texttt{gemini-3.1-pro-preview}, whole PDF), on overlapping
non-error reports. The vendor step (modality $\to$ cross-tier $\to$ cross-vendor) is monotone and
small; all three independent vendors reproduce the headline binary at $\kappa\ge0.86$.
Registry-gated.}
\label{tab:xvendor}
\begin{tabular}{llrr}
\toprule
arm & role & headline-parser $\kappa$ & coarse $\kappa$ \\
\midrule
\texttt{gemini-3.1-pro-preview} (text) & modality control (same model) & 0.956 & 0.959 \\
\texttt{gemini-3.7-flash} (text) & cross-tier control (Gemini) & 0.948 & 0.947 \\
Qwen3.6-27B & \textbf{cross-vendor} & \textbf{0.888} & 0.896 \\
DeepSeek V4 Flash & \textbf{cross-vendor} & \textbf{0.884} & 0.895 \\
MiniMax-M2.7 & \textbf{cross-vendor} & \textbf{0.857} & 0.857 \\
\midrule
\multicolumn{2}{l}{DeepSeek vs.\ Qwen (two non-Google vendors)} & 0.940 & 0.958 \\
\bottomrule
\end{tabular}
\end{table}

Numeric agreement on the value itself is $82$--$89\%$ within $5\%$ between independent vendors.
The residual disagreement is directional and concentrated: $20$--$23$ of the flips per vendor are
the primary counting a per-m$^2$ figure that the text-only vendor does not, and about $60\%$ of
those are multi-intensity reports (a per-area figure disclosed alongside a revenue- or
employee-normalised one), where which intensity is extracted is genuinely ambiguous --- the class
the pre-registered gold-set adjudication is designed to settle. Cross-vendor agreement is a
reproducibility bound that removes the shared-provider confound; it is not, on its own, a validity
claim, which continues to rest on literal grounding and the recall audit.

\section{Implementation details of the scoring and extraction pipelines}\label{app:implementation}

\paragraph{Credibility scoring (P1--P2)} Each report is presented to
the scoring model as three rendered page images: the cover page plus
two body pages sampled with a per-document deterministic seed, at
800\,px maximum dimension. The extracted document text accompanies the
images, truncated to 4{,}000 characters. The system instruction is: \emph{``You are a
multi-dimensional ESG-disclosure rater. Score the report on 5
dimensions using the rubric.''} The full SSSR rubric (dimension
definitions, sub-score formulas, and 0--3 anchor scales) is injected
into the prompt, and the model is instructed to return a single JSON
object with the five dimension scores and a one-sentence note per
dimension. Decoding uses temperature 0.1 with JSON-constrained output
and up to five retries with exponential backoff. All validators in
Table~\ref{tab:models} use the identical prompt and configuration;
Gemma validators receive the text channel only.

\paragraph{Disclosure-fitness extraction (P4)} The census was
extracted in two arms under an identical protocol: an initial
200-report arm drawn within region strata most-recent-reporting-year
first (a recency selection with no inclusion probabilities, which is
why no design weights exist), and the 302-report remainder extracted
subsequently to complete the frame. The extractor
(\texttt{gemini-3.1-pro-preview}) uploads
the \emph{entire} report PDF (no page sampling or truncation) and
requests a strict JSON object with ten fields (reporting year, absolute
Scope~1+2, GHG intensity value and unit, energy intensity value and
unit, gross floor area, property type, country or region, and an
evidence note). The instruction reads, verbatim: \emph{``Extract ONLY
values explicitly disclosed in this document. If a value is not
disclosed, return null --- never guess or infer a number.''} Decoding
parameters are the provider defaults (no temperature override), which
we note for transparency; categorical and numeric cross-extractor
agreement is reported in Section~\ref{sec:readiness} and
\ref{app:crrem}. The downstream class assignment is fully
deterministic: a rule-based unit parser normalises intensities to
kgCO$_2$e$\cdot$m$^{-2}\cdot$yr$^{-1}$, applies a physical-plausibility
band of $[1, 2000)$, and assigns the classes of Table~\ref{tab:gap}
by regular-expression matching on the disclosed denominators, with a
residual \emph{other-intensity} class so parser gaps are reported
rather than absorbed into a substantive class. Raw extraction outputs
are immutable; the three documented corrections (each with its
evidential basis) live in a released corrections file applied at load.
Every stage definition is fixed in the released measurement
specification, every number in this paper is regenerated from the raw
extracts by the released claims registry (exit-non-zero on drift), and
the unit parser is pinned by a frozen 400-string fixture with a
mutation test.

\paragraph{Identity anonymisation (leakage probe)} Text-channel
anonymisation is automated: the firm's ticker and every token of length
$\geq 4$ from its registered names (excluding legal-suffix stopwords
such as ``Inc'', ``Ltd'', ``Group'') are replaced, case-insensitively,
with the placeholder \texttt{[CO]}. The vision-channel arm omits the
cover page (where logos, branded chart styles and recognisable
building photography concentrate) and samples body pages only, with a
deterministic per-document seed.

\paragraph{Release} The verbatim prompts, the full rubric text, all
model identifiers and configurations, and the extraction and
classification code are included in the repository described under
Data availability, so every pipeline stage can be re-run end to end.

\clearpage
\bibliographystyle{elsarticle-num}
\bibliography{refs}

\end{document}